\documentclass[12pt]{article}
\usepackage{graphicx}
\usepackage{aas_macros}

\usepackage[margin=1in]{geometry}

\usepackage{amssymb}
\usepackage{amsmath}

\usepackage{xspace}
\usepackage{caption} 

\title{\bf A diffuse core in Saturn \\revealed by ring seismology}
\date{}

\author{Christopher R. Mankovich$^{1*}$ \& Jim Fuller$^2$}

\begin{document}


\maketitle

\begin{enumerate}
 \item {Division of Geological and Planetary Sciences, Mailcode 150-21, California Institute of Technology, Pasadena, CA 91125, USA}
 \item {TAPIR, Mailcode 350-17, California Institute of Technology, Pasadena, CA 91125, USA}
\end{enumerate}

\textbf{
The best constraints on the internal structures of giant planets have historically come from measurements of their gravity fields\cite{2018Natur.555..223K,2018Natur.555..227G,2019Sci...364.2965I}.
These gravity data are inherently mostly sensitive to a planet's outer regions, providing only loose constraints on the deep interiors of Jupiter\cite{2018Natur.555..227G,2017GeoRL..44.4649W,2017A&A...606A.139N} and Saturn\cite{2019ApJ...879...78M,2020ApJ...891..109M}.
This fundamental limitation stymies efforts to measure the mass and compactness of these planets' cores, crucial properties for understanding their formation pathways and evolution\cite{2017ApJ...840L...4H, 2018oeps.book..175H}.
However, studies of Saturn's rings have revealed waves driven by pulsation modes within Saturn\cite{2013AJ....146...12H,2014MNRAS.444.1369H,2019Icar..319..599F,2019AJ....157...18H}, offering independent seismic probes of Saturn's interior\cite{1993Icar..106..508M,2014Icar..242..283F,2019ApJ...871....1M}.
The observations reveal gravity mode (g~mode) pulsations which indicate that a part of Saturn's interior is stably stratified by composition gradients, and the g~mode frequencies directly probe the buoyancy frequency within the planet\cite{2014Icar..242..283F}.
Here, we compare structural models with gravity and seismic measurements to show that the data can only be explained by a diffuse, stably stratified core-envelope transition region in Saturn extending to {approximately $60\%$ of the planet's radius and containing approximately 17 Earth masses of ice and rock.}
The gradual distribution of heavy elements constrains mixing processes at work in Saturn, and it may reflect the planet's primordial structure and accretion history.
}

Measurements of Jupiter's even zonal gravity harmonics $J_{2n}$ ($n=1,\, 2,\,\ldots$) by the Juno spacecraft have raised the possibility of a gradual core-envelope transition within Jupiter\cite{2017GeoRL..44.4649W,2020AREPS..48..465S}.
At Saturn, the gravity field measured by the Cassini spacecraft is complicated by the large contribution from deep zonal flows\cite{2019Sci...364.2965I,2019GeoRL..46..616G,2021MNRAS.501.2352G}, and disentangling these dynamical contributions from those of the rigidly rotating deep interior remains an outstanding challenge for understanding Saturn's deep structure.
While it is known that Saturn's low-degree gravity harmonics require some form of central density enhancement in Saturn\cite{2005AREPS..33..493G}, it is unknown to what extent this enhancement takes the form of a compact core versus a diffuse core structure akin to the one proposed for Jupiter.
Furthermore, gravity data offer no direct information about the phase (fluid vs. solid) or stratification (mixed by convection vs. compositional layering) of the planet's interior, leaving open the question of whether the interiors of the gas giants are fully convective as in the conventional picture\cite{1968ApJ...152..745H}.

A unique opportunity to answer these questions in the case of Saturn comes from studies of its rings.
Stellar occultation experiments in Saturn's C ring carried out by Cassini have revealed a Saturn pulsation spectrum\cite{2013AJ....146...12H,2014MNRAS.444.1369H,2016Icar..279...62F,2019Icar..319..599F,2019AJ....157...18H} dominated by fundamental modes (f~modes)\cite{1991Icar...94..420M,1993Icar..106..508M,2019ApJ...871....1M,2020AGUA....100142M}, but also enriched by internal gravity waves (g~modes) trapped closer to the planet's center\cite{2014Icar..242..283F}.
Because gravity waves are restored by buoyancy, the presence of g~modes in Saturn implies that part of the interior is stabilized against convection by a composition gradient.
{Such composition gradients are neglected in conventional layered structure models for gas giants, which instead assume a small number of distinct, chemically homogeneous layers, typically with a discrete core of heavy elements at the center\cite{1999P&SS...47.1201G,1999P&SS...47.1183G,2004ApJ...609.1170S,2013Icar..225..548N,2016ApJ...820...80H}.}
{Increasingly, theoretical work has seriously considered the role of stable heavy element gradients or helium gradients within Jupiter or Saturn\cite{2012A&A...540A..20L,2013NatGe...6..347L,2015MNRAS.447.3422N,2016ApJ...829..118V,2019ApJ...872..100D,2020ApJ...891..109M}, and to date ring seismology at Saturn provides the sole means of probing these gradients directly.}
{The present work is motivated by the recent characterization\cite{2019Icar..319..599F} of a previously detected inner C ring wave\cite{2011Icar..216..292B} (W76.44) that appears to be connected to Saturn's g~modes but has an unexpectedly low frequency relative to the published model\cite{2014Icar..242..283F}.}
{We find here that this pattern constitutes a major new constraint on Saturn's deep composition gradient, its low frequency pointing toward a more extended gradient than previously thought.}

The normal modes of a slowly rotating planet can be indexed by the radial order $n$ characterizing their radial structure, and by the spherical harmonic degree $l$ and azimuthal order $m$ characterizing their angular structure. The rapid rotation of Saturn couples modes with different values of $l$, producing a complex spectrum (see Methods). Nonetheless, each normal mode is typically dominated by a single ($n , l, m)$ component, and the value of $m$ can be measured directly from the ring wave's azimuthal structure.
Of particular value for probing Saturn's deep interior are the $m=-2$ g~modes, which provide stringent constraints on the extent of Saturn's stably stratified core-envelope transition region.
Here we use the gravity data and ring seismology constraints to jointly quantify the extent of Saturn's interior stable stratification.
To this end, we compute oblate Saturn interior models and calculate the spectrum of normal mode oscillations for each model, in addition to the zonal gravity harmonics $J_2$, $J_4${, and} $J_6$.
The latter are calculated perturbatively using a fourth-order theory of figures to solve for the consistent shape and background potential for each oblate, rotating model.
We consider several {structural} parameterizations whose common feature is a core-envelope transition region over which the heavy element mass fraction $Z$ and helium mass fraction $Y$ vary continuously.
The salient parameters are the shallow and deep heavy element mass fractions \ensuremath{Z_{\rm out}}\xspace and \ensuremath{Z_{\rm in}}\xspace and the {outer radius $r_{\rm out}$} of the {composition gradient} connecting the two.
{We find that a helium gradient consistent with hydrogen-helium phase separation\cite{2018PhRvL.120k5703S} is also required to fit the gravity data and contributes significantly to the buoyancy, but it is not required by the seismology alone (see Methods).}

\begin{figure}
  \centering
  \includegraphics[width=\textwidth]{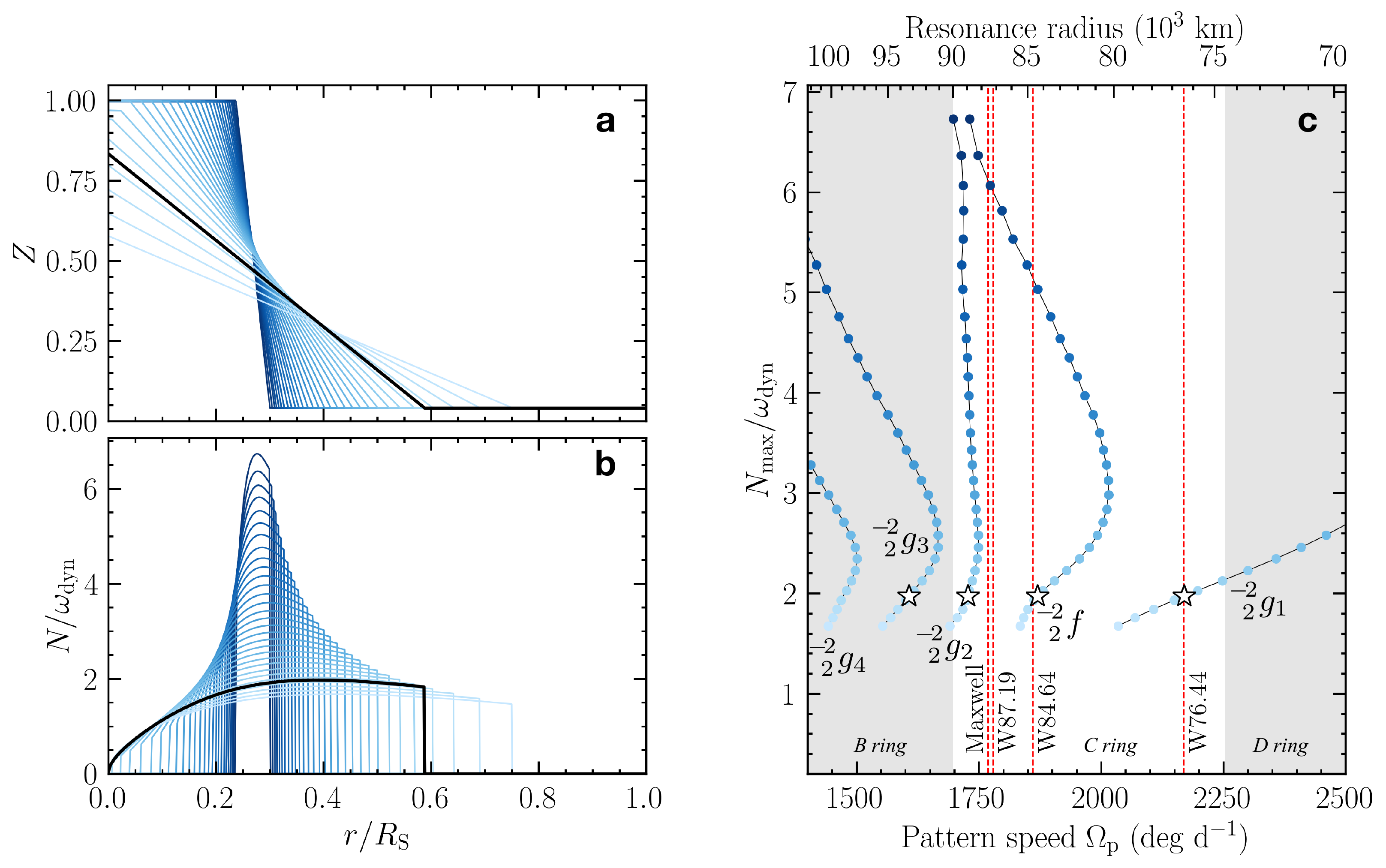}
  \caption{\label{fig.cavity_width}
  {\bf Ring seismology constrains Saturn's core-envelope interface.}
  {\bf a,} Candidate distributions of heavy elements as a function of radius within Saturn.
  {\bf b,} The resulting Brunt-V\"ais\"al\"a frequency $N$, scaled to Saturn's natural frequency $\omega_{\rm dyn}=(GM_{\rm S}/R_{\rm S}^3)^{1/2}$.
  The stably stratified g~mode cavity is the region where $N>0$.
  {\bf c,} Modeled Saturn $m=-2$ mode frequencies compared to observed ring waves.
  Frequencies are plotted as a function of the maximum value of $N$ attained in the corresponding interior model.
  The pattern speed (bottom axis) is the frequency observable in an inertial reference frame.
  The top axis shows the corresponding Lindblad resonance radius in Saturn's ring plane, measured from Saturn's center.
  Vertical dashed lines mark the locations of the $m=-2$ spiral density waves observed at outer Lindblad resonances in the C ring.
  The model highlighted with bold curves in {\bf a}-{\bf b} and star symbols in {\bf c} is the best model from our joint gravity-seismology fits (Figs.~\ref{fig.profiles}-\ref{fig.ompat_j2n}).
  }
\end{figure}

Fig.~\ref{fig.cavity_width} compares our models' mode frequencies to the ring data.
Saturn's g~mode pulsation spectrum is controlled principally by the width of the stable density stratification between the ice/rock-rich core and the hydrogen-dominated envelope (Fig.~\ref{fig.cavity_width}a-b).
Fig.~\ref{fig.cavity_width}c highlights the special utility of {the wave W76.44} for constraining the extent of Saturn's stable stratification.
We identify this wave with Saturn's lowest radial order $l=2$, $m=-2$ g~mode (\ensuremath{_{\,\,\,\,2}^{-2}g_1}\xspace), which takes the form of a quasi-interface mode partially trapped on the diffuse core-envelope interface (Extended Data Figs.~\ref{fig.eigs_best}-\ref{fig.eigs_cavity_width}).
For sharp core-envelope transitions the Brunt-V\"ais\"al\"a frequency $N$ becomes large and this mode becomes increasingly trapped on the narrow interface, obtaining small surface gravity perturbations and frequencies too high to resonate in the C ring.
In the opposite extreme of gradual core-envelope transitions extending toward Saturn's full radius, this mode obtains larger gravity perturbations but lower frequencies, moving its Lindblad resonance into the middle C ring and leaving W76.44 unexplained.
The observed frequency instead strongly favors an intermediate level of stratification, corresponding to an extended core-envelope transition {out to $r_{\rm out}\approx0.6\,R_{\rm S}$.}
This conclusion is bolstered by the remainder of the known $m=-2$ patterns: the same interface width favored by W76.44 produces the best simultaneous fit to W84.64 (via the mode \ensuremath{_{\,\,\,\,2}^{-2}f}\xspace) and W87.19/Maxwell (via \ensuremath{_{\,\,\,\,2}^{-2}g_2}\xspace). The close pair of distinct waves W87.19 and Maxwell is most likely generated by an avoided crossing between \ensuremath{_{\,\,\,\,2}^{-2}g_2}\xspace and an $l\gg2$ g~mode that is not accurately captured by our perturbative treatment of rotation (see Methods).

\begin{figure}
  \centering
  \includegraphics[width=0.6\textwidth]{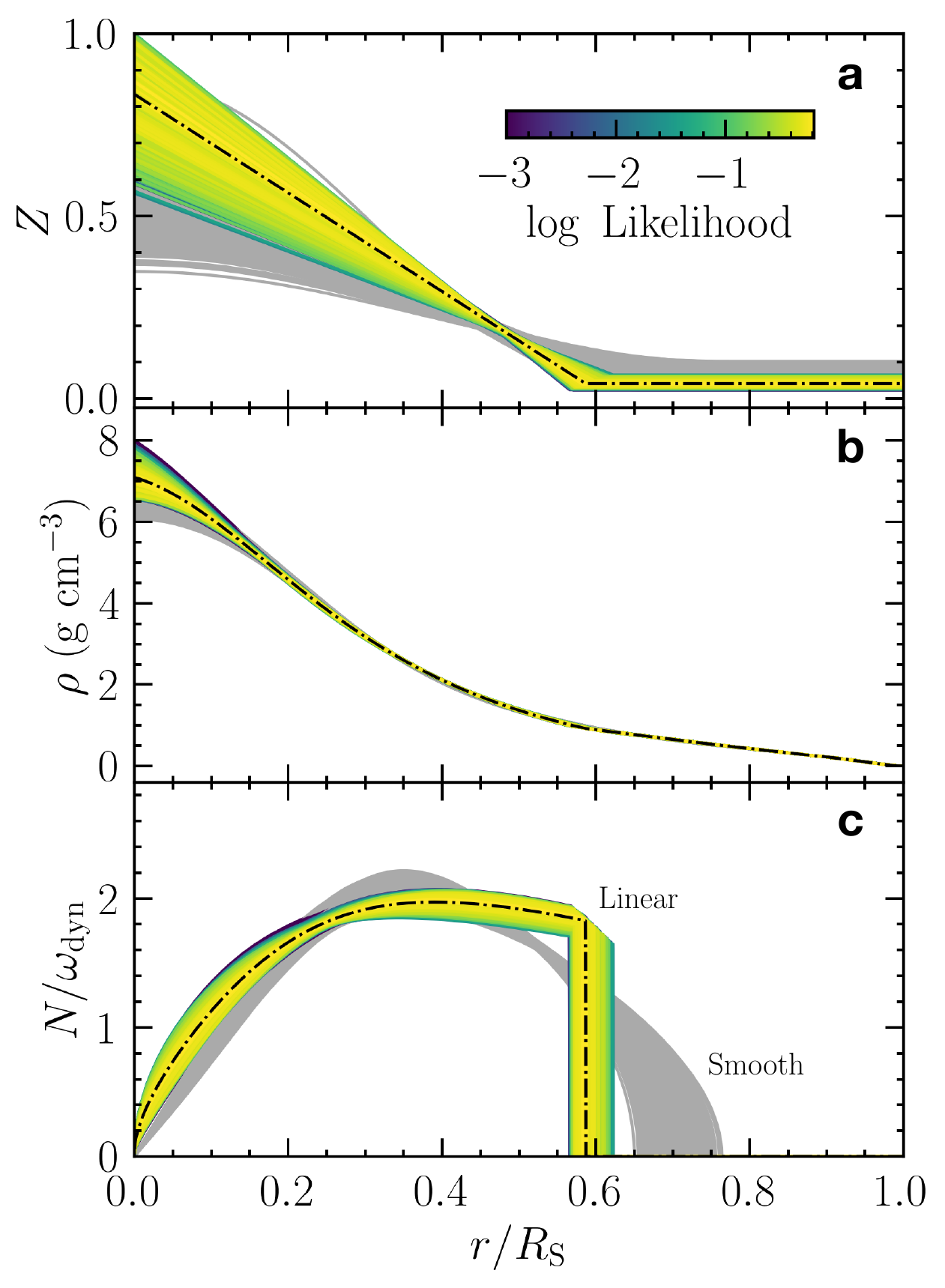}
  \caption{\label{fig.profiles}
  {\bf Saturn's internal structure deduced from ring seismology and the gravity field.}
  Profiles of heavy element mass fraction ({\bf a}), mass density ({\bf b}), and Brunt-V\"ais\"al\"a frequency ({\bf c}) are shown as functions of mean level radius for {1024} randomly chosen models from our joint gravity-seismology sample.
  Each profile is colored by the corresponding model's log likelihood (see Methods).
  The single best model in the sample is overlaid in dot-dashed black curves.
  Grey curves in the background {impose smooth composition profiles instead of linear profiles} (see Methods).
  }
\end{figure}

To quantitatively constrain Saturn's interior structure and composition profile, we assign any proposed model a likelihood based on its ability to reproduce Saturn's low-degree zonal gravity harmonics {$J_2$, $J_4$, and $J_6$} along with the frequencies of 3 of the 4 observed $m=-2$ ring patterns, and generate statistical samples of models using Markov chain Monte Carlo (see Methods).
Fig.~\ref{fig.profiles} summarizes the baseline distribution of Saturn internal structures that we jointly retrieve from gravity data and ring seismology. As the cursory exploration in Fig.~\ref{fig.cavity_width} suggested, the ring seismology constrains Saturn's core-envelope transition precisely (Fig.~\ref{fig.profiles}c), yielding transition widths $r_{\rm out}/R_{\rm S}=0.59\pm0.01$ (mean and standard deviation).
This finding is remarkably insensitive to {variations in Saturn's helium distribution, bulk water ice to rock ratio, 1-bar temperature, and superadiabatic thermal stratification} (Extended Data Table~\ref{tab.models}).

{The joint seismology/gravity fit also tightly constrains Saturn's density profile (Fig.~\ref{fig.profiles}b), although the exact composition profile is less clear, due to the degenerate roles that the ice to rock mixing ratio and the overall heavy element mass fraction play in determining the density.
Our best-fitting models predict central metallicities of typically $\ensuremath{Z_{\rm in}}\xspace=0.8$, suggesting that some hydrogen and helium are mixed down to the planet's center, but equally likely models with $\ensuremath{Z_{\rm in}}\xspace\approx1$ are achieved when the heavy element mass is dominated by water ice rather than rock (Extended Data Fig~\ref{fig.z2-fice}).}
The derived envelope metallicities {$\ensuremath{Z_{\rm out}}\xspace=0.041\pm0.009$} match the $\approx \! 3\times$ solar enrichment of NH$_3$ measured in thermal emission at Saturn's equator\cite{2011Icar..214..510F}, but they fall short of the $\approx \! 9\times$ solar CH$_4$ abundance thought to be representative of Saturn's deep atmospheric C abundance\cite{2009Icar..199..351F}. This known tension with interior models constrained by the gravity field\cite{2013Icar..225..548N,2013ApJ...767..113H,2019ApJ...879...78M} may indicate a more complicated envelope structure in Saturn than has been considered to date. We note that when fit independently of $J_{2n}$, the ring seismology is compatible with envelope metallicities ranging from sub-solar to $\gtrsim \! 9\times$ solar (Extended Data Table~\ref{tab.models}).

\begin{figure}
  \centering
  \includegraphics[width=0.8\textwidth]{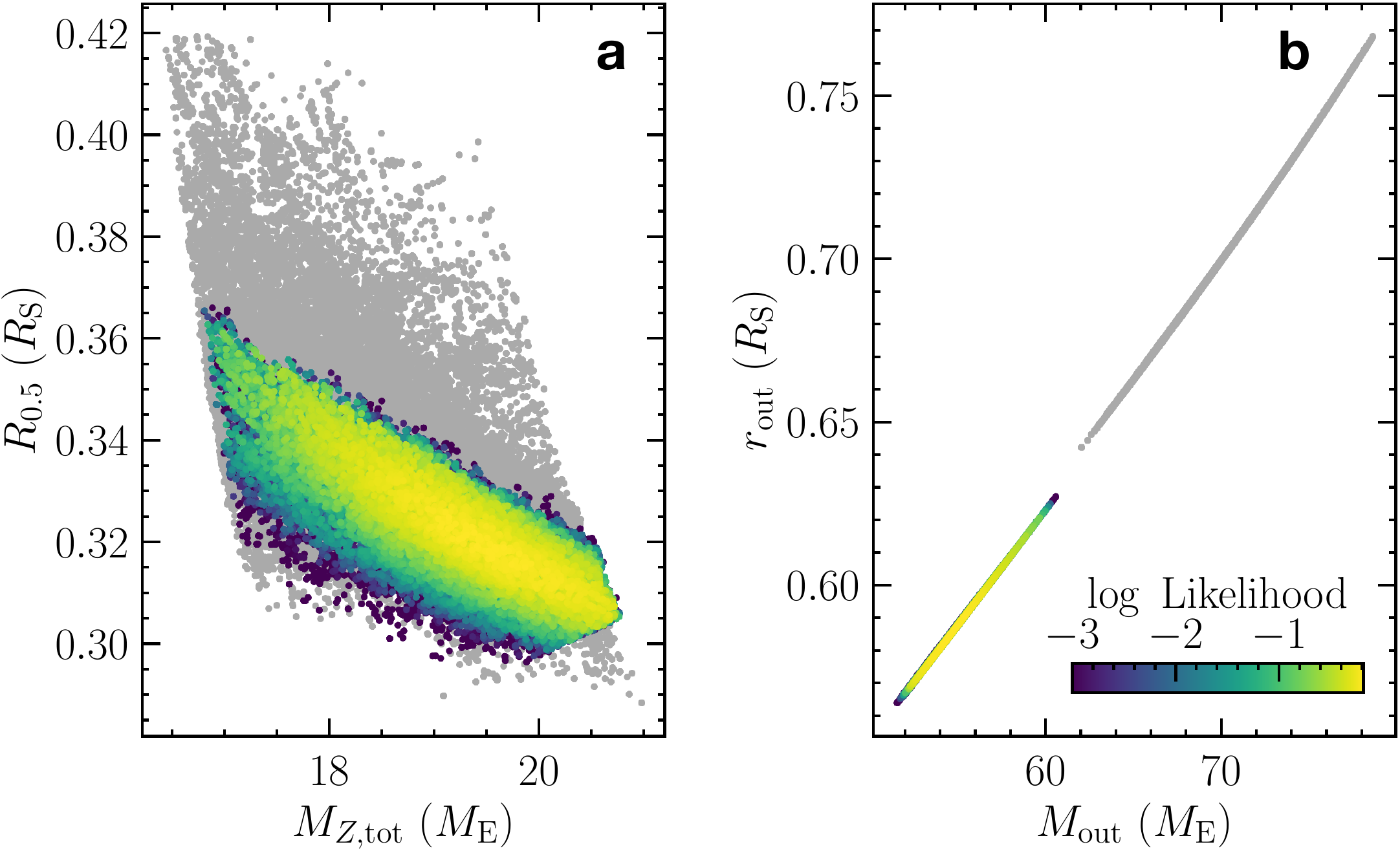}
  \caption{\label{fig.core_proxy}
  {\bf Saturn's ``core mass'' and ``core radius'' from gravity and ring seismology.}
{
    {\bf a}, the radius containing half the model's total heavy element mass, as a function of the model's total heavy element mass.
    {\bf b}, the outer radius of the stably stratified region as a function of its mass.
    Our baseline model is colored by log likelihood; the grey distributions impose smooth composition profiles instead of linear profiles (see Methods).
    }
  }
\end{figure}

{Our models place tight constraints on the mass and size of the heavy element core of Saturn, even as the dilute nature of this core requires a more nuanced description than in traditional layered models.
The \emph{total} heavy element mass in the models is $M_{Z,{\rm tot}} = 19.1\pm1.0$ Earth masses, consistent with estimates from layered models\cite{1999P&SS...47.1183G,2004ApJ...609.1170S,2013Icar..225..548N,2019ApJ...879...78M}.
Defining the extended core as the region containing half of the model's total heavy element mass,
Fig.~\ref{fig.core_proxy} shows that the core region occupies an effective radius of $R_{0.5} = 0.32\pm0.01$ Saturn radii.
If we alternatively define the core boundary as the surface of the stable region at $r_{\rm out} = 0.59 \pm 0.01$ Saturn radii, we infer a core mass of $M_{\rm stab} = 55.1 \pm 1.7$ Earth masses, of which $M_{Z,{\rm stab}} = 17.4 \pm 1.2$ Earth masses is rock and ice.
Independent of the somewhat arbitrary core definition,} we caution that these uncertainties are likely underestimated because they do not include systematic uncertainties inherent to our choice of parameterization (see Methods {for a model selection analysis}).
Nonetheless, due to the extended nature of {this stratification}, these core masses and radii are substantially larger than prior estimates\cite{1999P&SS...47.1183G,2004ApJ...609.1170S,2013Icar..225..548N,2019ApJ...879...78M} based on gravity data and simple {layered} models with pure {heavy element} cores.
{We finally note that in our most likely models, the composition gradient extends all the way to Saturn's center. Models with a chemically homogeneous inner core are somewhat less likely but cannot be ruled out (see Methods).}

\begin{figure}
  \centering
  \includegraphics[width=\textwidth]{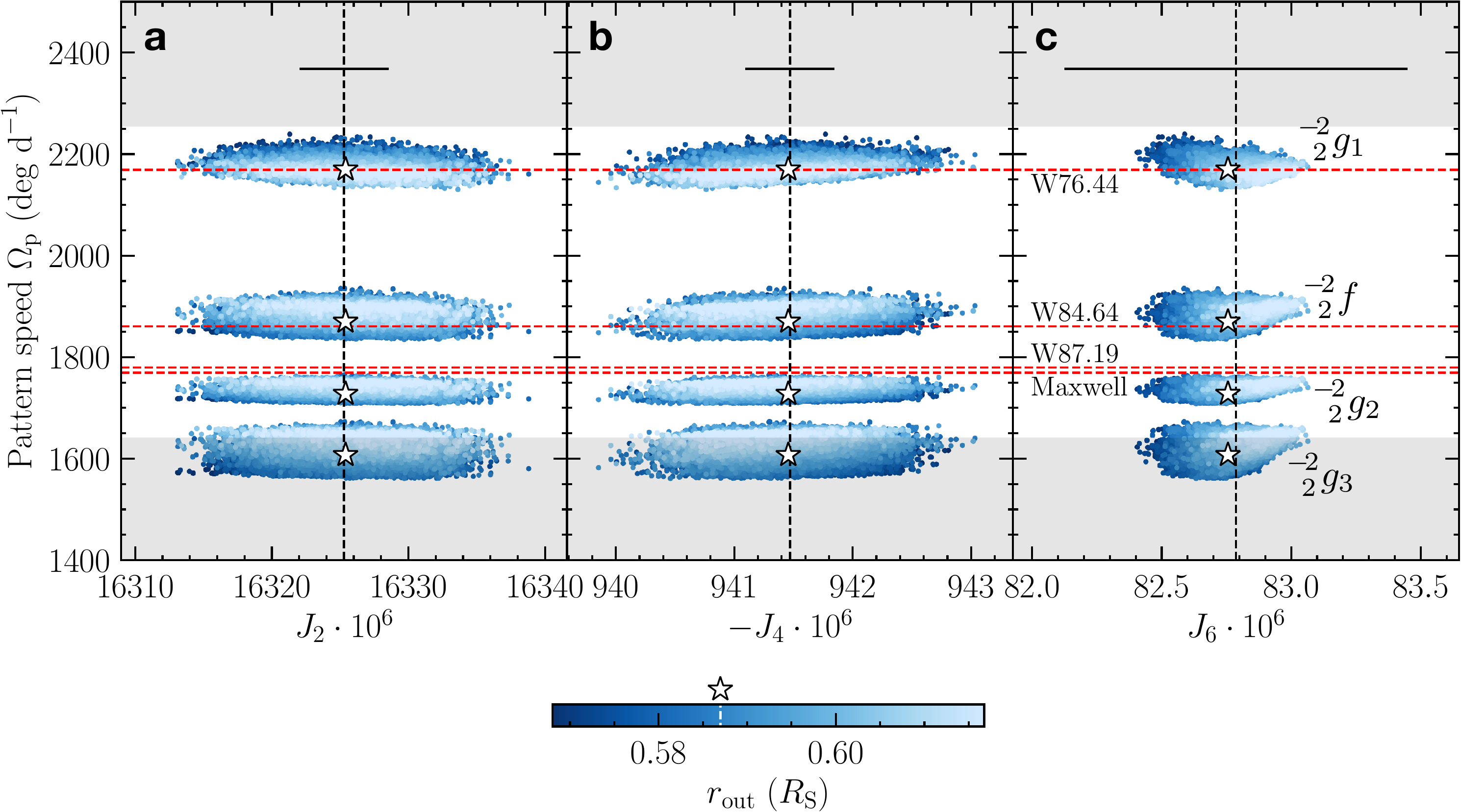}
  \caption{\label{fig.ompat_j2n}
  {\bf Saturn $m=-2$ mode spectra and gravity harmonics.}
  Mode pattern speeds $\Omega_{\rm p}$ are shown as a function of the gravity harmonics $J_2$ ({\bf a}), $J_4$ ({\bf b}), and $J_6$ ({\bf c}) for all models in our baseline sample. Each interior model is colored by the radial width $w$ of its g~mode cavity.
  Observed pattern speeds of $m=-2$ spiral density waves in the C ring are plotted as horizontal dashed lines.
  The vertical dashed lines indicate the $J_{2n}$ measured by Cassini{, with dynamical contributions subtracted to yield the rigid-body harmonics (see Methods)}.
  The black horizontal bars represent the effective $1\sigma$ uncertainty in gravity harmonics given our numerical method.
  Stars indicate the single most likely interior model highlighted in Figs.~\ref{fig.cavity_width}-\ref{fig.profiles}.
  }
\end{figure}

Even as the seismology stringently constrains interior structure, an inspection of the same distribution of models against the observed wave frequencies and gravity harmonics (Fig.~\ref{fig.ompat_j2n}) shows that none of the models are entirely satisfactory.
As in Fig.~\ref{fig.cavity_width}, these calculations do not reproduce the fine frequency splitting between the Maxwell and W87.19 density waves{, but this and the similar fine splitting prominent in the $m=-3$ spectrum can be understood in terms of interactions with high-degree g~modes (see Methods).}
Robustly explaining this fine structure may require a full treatment of latitudinal differential rotation {or a non-perturbative treatment of rotation, both of} which tend to strengthen coupling between modes of different $l$\cite{2014Icar..242..283F}.

The second shortfall of the models seen in Fig.~\ref{fig.ompat_j2n} is that they systematically underestimate the frequency of W87.19 (or the Maxwell wave) predicted by the \ensuremath{_{\,\,\,\,2}^{-2}g_2}\xspace mode{.}
This may indicate that the shape of Saturn's $Z$ distribution differs from the simple functional form imposed here, leading the models to overestimate Saturn's true g~mode period spacing (see Methods) even as the width of the core-envelope transition region is effectively anchored by the frequency of W76.44 (Fig.~\ref{fig.cavity_width}c).
A related issue is that the frequencies of \ensuremath{_{\,\,\,\,2}^{-2}g_2}\xspace and \ensuremath{_{\,\,\,\,2}^{-2}f}\xspace are effectively repelled from each other as a result of an avoided crossing (Fig.~\ref{fig.cavity_width}c), an effect enhanced by the substantial overlap between the f~mode and g~mode cavities.
The proximity of the corresponding observed frequencies suggests that these two modes are more weakly coupled in reality, again indicating that {more complicated} functional forms for $Z(r)$ and $Y(r)$ may be necessary to capture the details of Saturn's deep composition stratification.

Our finding of a diffuse core in Saturn stands in contrast with the more centrally condensed heavy element {distributions} predicted by classic core accretion models\cite{1996Icar..124...62P}, {instead underscoring the more gradual distributions predicted by recent accretion models\cite{2017ApJ...836..227L,2017ApJ...840L...4H,2020arXiv201014213O}.}
{Indeed, evolutionary models suggest that Saturn's inner half by radius likely retains its stratification even as convection driven by cooling rapidly homogenizes the outer regions\cite{2016ApJ...829..118V,2020A&A...638A.121M}.}

Alternatively, Saturn may have formed with a more abrupt core/envelope interface that eroded over time into the diffuse structure seen today, a hypothesis allowed by the miscibility of water, silicates, and iron in fluid metallic hydrogen \cite{2012ApJ...745...54W,2012PhRvL.108k1101W,2013ApJ...773...95W}. In this scenario, the upward mixing of dense elements from the core boundary may be limited by layered double-diffusive convection\cite{2017ApJ...849...24M}, and it remains unclear whether a compositional stratification out to $0.6$ Saturn radii could be achieved. A third and more remote possibility is that the diffuse core was established late by a head-on massive impact\cite{2019Natur.572..355L}.

Our results for Saturn echo the dilute core argued to exist in Jupiter\cite{2017GeoRL..44.4649W,2019ApJ...872..100D} on the basis of its gravity field.
But unlike {the static gravity field}, seismology directly probes fluid stability.
The ring seismology requires that Saturn's interior be convectively stable over approximately half the planet's radius, a fundamental departure from published interior models constrained by the gravity field\cite{2019ApJ...879...78M,2013Icar..225..548N,2013ApJ...767..113H}.
Over this stable region, heat is likely transported by double-diffusive mixing, possibly in the form of layered convection\cite{2012ApJ...750...61M,2013ApJ...768..157W}.
This large stably stratified region would delay the planet's cooling, helping to explain Saturn's surprisingly high luminosity\cite{2013NatGe...6..347L}.
The more weakly constrained dilute core structure in Jupiter may similarly take the form of a continuous heavy element gradient supporting a stable stratification and g~modes.
{Finally, Saturn's stably stratified region subsumes much of the electrically conductive part of Saturn's interior, posing an ostensible challenge for explaining Saturn's magnetic dynamo and its axisymmetric external field.
Nonetheless our model's convective envelope reaches pressures exceeding $10^6\ {\rm bar}$ where a large electrical conductivity of order $10^4\ {\rm S\ m}^{-1}$ is expected\cite{2014Icar..239..260L} and hence our proposed structure remains conducive to shell dynamo action.
More complex internal structures, perhaps containing two or more stably stratified shells, may be needed for a complete picture\cite{1982GApFD..21..113S,2010GeoRL..37.5201S,2020Icar..34413541C}.}

\clearpage
\section*{Methods}

{\bf Planetary interior models.}
We compute rigidly rotating models satisfying Saturn's total mass and equatorial radius. We solve for the self-consistent potential and hydrostatic shape using a 4th-order theory of figures\cite{1978ppi..book.....Z,2017A&A...606A.139N} (ToF), yielding the even zonal gravity harmonics $J_2$, $J_4$, $J_6$, and $J_8$.
Our baseline bulk rotation period is 10h33m38s as derived from ring seismology\cite{2019ApJ...871....1M}.
We denote by $r$ the volume-equivalent spherical radius of equipotential surfaces in ToF.
The models apply the Militzer \& Hubbard\cite{2013ApJ...774..148M} \textit{ab initio} equation of state (EOS) for H and He, combined with the Saumon-Chabrier-van Horn EOS\cite{1995ApJS...99..713S} for He under an additive-volume law to model arbitrary mixtures of H and He\cite{2016A&A...596A.114M}.
Heavy elements are treated as a mixture of silicates and water ice, represented by ANEOS\cite{aneos} serpentine and water ice respectively, {with the mass mixing ratio \ensuremath{f_{\rm ice}}\xspace of the two components treated as a free parameter.}
{The profile of the heavy element mass fraction $Z(r)$ is constructed to connect a compositionally uniform outer envelope to a compositionally uniform inner core:
\begin{equation}
\label{eq.z_profile_linear}
  Z(r)=\left\{\begin{array}{ll}
    \ensuremath{Z_{\rm out}}\xspace, & r >= \ensuremath{r_{\rm out}}\xspace, \\
    \displaystyle \ensuremath{Z_{\rm out}}\xspace + (\ensuremath{Z_{\rm in}}\xspace-\ensuremath{Z_{\rm out}}\xspace)\left(\frac{\ensuremath{r_{\rm out}}\xspace - r}{\ensuremath{r_{\rm out}}\xspace - \ensuremath{r_{\rm in}}\xspace}\right), & \ensuremath{r_{\rm in}}\xspace < r < \ensuremath{r_{\rm out}}\xspace, \\
    \ensuremath{Z_{\rm in}}\xspace, & r <= \ensuremath{r_{\rm in}}\xspace.
  \end{array}\right.
\end{equation}
We alternatively consider smooth profiles that take a sigmoid shape in the gradient region:
\begin{equation}
\label{eq.z_profile_sigmoid}
  Z(r)=\ensuremath{Z_{\rm out}}\xspace + (\ensuremath{Z_{\rm in}}\xspace-\ensuremath{Z_{\rm out}}\xspace)\sin^2 \left[\displaystyle \frac{\pi}{2}\left(\frac{\ensuremath{r_{\rm out}}\xspace-r}{\ensuremath{r_{\rm out}}\xspace-\ensuremath{r_{\rm in}}\xspace}\right) \right], \quad \ensuremath{r_{\rm in}}\xspace < r < \ensuremath{r_{\rm out}}\xspace.
\end{equation}
Here the outer envelope metallicity $\ensuremath{Z_{\rm out}}\xspace$ and the inner/outer transition radii $\ensuremath{r_{\rm in}}\xspace$ and $\ensuremath{r_{\rm out}}\xspace$ are free parameters.
With these fixed, the inner core metallicity $\ensuremath{Z_{\rm in}}\xspace$} modulates the mean density and thus must be adjusted during ToF iterations to satisfy Saturn's equatorial radius $60,268\ {\rm km}$ at $P=1\ {\rm bar}$\cite{1985AJ.....90.1136L}.
{Our baseline four-parameter models fix $\ensuremath{r_{\rm in}}\xspace=0$; alternative samples that allow \ensuremath{r_{\rm in}}\xspace to vary freely are discussed below.}

Another parameter $T_1$ specifies the temperature of the model's outer boundary at $P=1\ {\rm bar}$, establishing the interior entropy. Our baseline sample fixes $T_1=135~{\rm K}$ following Voyager radio occultation measurements\cite{1985AJ.....90.1136L,1992AJ....103..967L} {and assumes that
deeper temperatures follow an adiabatic thermal stratification}. In reality, an otherwise convective environment that is stabilized by composition gradients {generally attains} a superadiabatic stratification in temperature.
We calculate $N^2$ in the practical form\cite{1991ApJ...367..601B}
\begin{equation}
\label{eq.brunt}
  N^2=\frac{g^2\rho}{P}\frac{\chi_T}{\chi_\rho}\left(\nabla_{\rm ad}-\nabla+B\right).
\end{equation}
where $\nabla\equiv d\ln T/d\ln P$, $\nabla_{\rm ad}\equiv (\partial\ln T/\partial\ln P)_s$, and
\begin{equation}
\label{eq.define_brunt_b}
B\equiv-\frac{1}{\chi_T}\left(\chi_Y\frac{d\ln Y}{d\ln P}+\chi_Z\frac{d\ln Z}{d\ln P}\right)
\end{equation}
accounts for the gradients in composition. These relations make use of the definitions
\begin{equation}
\label{eq.define_chirho_chit_chiy_chiz}
\chi_T\equiv\left(\frac{\partial\ln P}{\partial\ln T}\right)_{\rho, Y, Z},\quad
\chi_\rho\equiv\left(\frac{\partial\ln P}{\partial\ln \rho}\right)_{T, Y, Z},\quad
\chi_Y\equiv\left(\frac{\partial\ln P}{\partial\ln Y}\right)_{\rho, T, Z},\quad
\chi_Z\equiv\left(\frac{\partial\ln P}{\partial\ln Z}\right)_{\rho, T, Y}.
\end{equation}
From the perspective of the buoyancy frequency (\ref{eq.brunt}), a superadiabatic thermal stratification $\nabla>\nabla_{\rm ad}$ tends to mitigate a stabilizing composition gradient $B>0$.
A basic degeneracy therefore affects the interpretation of the $N(r)$ profiles that we infer (Fig.~\ref{fig.profiles}c){: we} cannot rule out a superadiabatic thermal stratification if it is accompanied by a commensurately stronger composition contrast between core and envelope.

{We note however that a plausible few-fold increase in central temperature as a result of superadiabaticity within the composition gradient ($P\approx1$ to 10 Mbar) would imply $\nabla-\nabla_{\rm ad}\sim0.5$, a minor contribution to $N^2$ compared to the composition term, typically $B\approx6$ in our models.
Conversely, a perfectly compensating thermal gradient $\nabla-\nabla_{\rm ad}=B\approx6$ corresponding to the threshold of Ledoux instability would imply an impossible $10^6$-fold increase in central temperature.
Extended Data Fig.~\ref{fig.superad} presents models with superadiabatic thermal stratifications, taking $\nabla-\nabla_{\rm ad}$ as a constant in the gradient region for simplicity and sampling that quantity as an additional parameter.
Although these models' central temperatures vary by a factor of $\sim5$, the $N$ profiles are largely unchanged for the reasons just described. The central densities are also largely unchanged because the density is rather insensitive to temperature in a rock- and ice-dominated environment. We thus conclude that a superadiabatic thermal stratification in the core-envelope transition region is unlikely to modify our main findings.
}

Hydrogen and helium are expected to be immiscible in the fluid metallic H that dominates Saturn's interior deeper than $P\gtrsim10^6\ {\rm bar}$\cite{1973ApJ...181L..83S,1977ApJS...35..221S}, a prediction supported by \emph{ab initio} simulations\cite{2018PhRvL.120k5703S,2009PNAS..106.1324M,2013PhRvB..87q4105M,2009PhRvL.102k5701L,2011PhRvB..84w5109L} and infrared and radio data indicating a He-depleted atmosphere\cite{1980JGR....85.5871O,1984ApJ...282..807C,2000Icar..144..124C,2016Icar..276..141S,2018Icar..307..161K}.
Our model accordingly also includes a He gradient, assumed for simplicity to coincide spatially with the $Z$ gradient.
We take $\ensuremath{Y^\prime}\xspace\equiv Y/(1-Z)$, the He mass fraction relative to H+He, to follow a {linear profile exactly analogous Eq.~\ref{eq.z_profile_linear} but with \ensuremath{Y^\prime_{\rm out}}\xspace, \ensuremath{Y^\prime_{\rm in}}\xspace in place of \ensuremath{Z_{\rm out}}\xspace, \ensuremath{Z_{\rm in}}\xspace.
The alternative models with sigmoid $Z(r)$ (Eq.~\ref{eq.z_profile_sigmoid}) similarly assume a sigmoid profile for $\ensuremath{Y^\prime}\xspace(r)$.}
The deep He mass fraction \ensuremath{Y^\prime_{\rm in}}\xspace is a free parameter; values of approximately 95\% are predicted for the He-rich phase in Saturn's deep interior \cite{2016Icar..267..323P,2020ApJ...889...51M}. \ensuremath{Y^\prime_{\rm out}}\xspace is varied during ToF iterations to achieve a bulk He to H ratio $\langle{Y^\prime}\rangle=0.275$ representing that of the protosolar nebula\cite{2009ARA&A..47..481A}.
Although a \ensuremath{Y^\prime_{\rm in}}\xspace near unity implies a strong enrichment of He relative to H in the inner core, the H-He system overall is displaced by the large heavy element abundance there so that $Y$ itself is a non-monotone function of depth (Extended Data Fig.~\ref{fig.profiles_comparison}a-b).
{Nonetheless the tradeoff of H/He for heavier elements with increasing depth guarantees a positive buoyancy frequency throughout the gradient region.}

In no cases are uniform He distributions recovered: the data favor depletion of He from the outer envelope, consistent with thermal emission and occultation measurements\cite{1980JGR....85.5871O,1984ApJ...282..807C,2000Icar..144..124C,2016Icar..276..141S,2018Icar..307..161K}.
This is driven by gravity constraints: an alternative sample neglecting $J_{2n}$ {does produce models with a uniform He to H ratio that fit the ring seismology} (Extended Data Table 1).
We retrieve envelope mass ratios {$Y/(X+Y)=0.16\pm0.02$} (mean and standard deviation) for our baseline case, for an atmospheric He/H$_2$ mole ratio of {$0.084\pm0.009$}.
These abundances are compatible with constraints from Cassini UVIS/CIRS data\cite{2018Icar..307..161K}.
They exceed predictions from models that made a more detailed accounting of H-He immiscibility\cite{2015MNRAS.447.3422N,2016Icar..267..323P,2020ApJ...889...51M} but considered only simplistic heavy element distributions.

{In reality Saturn's deep rotation rate and 1-bar temperature are also not perfectly known, and thus we consider more general samples in which these parameters can vary.
We find that neither parameter can improve the quality of fit, and so neither bears strongly on the structure that we infer. The complete list of parameterizations pursued are summarized in Extended Data Table~\ref{tab.models} and described in `Parameter estimation' below.}

\noindent{\bf Theoretical mode spectra.}
Saturn's rapid rotation modifies its normal mode spectrum considerably as a result of Coriolis and centrifugal forces and the planet's oblateness.
We solve for normal modes using a method developed and described in detail in a prior study by Fuller\cite{2014Icar..242..283F}.
First we solve the linearized, adiabatic fluid perturbation equations in an angle-averaged formulation that separates solutions by spherical harmonic angular degree $l$. We term the solutions to the angle-averaged equations ``pseudo-modes."
These solutions account for rotation only through first-order Coriolis self-coupling; in the limit of slow rotation they recover the first-order frequency perturbation commonly applied in asteroseismology\cite{2013ARA&A..51..353C} that for each $l$ symmetrically splits the $m\neq0$ mode frequencies away from the $m=0$ frequency\cite{1951ApJ...114..373L}.
However, in contrast to the usual basis modes obtained in absence of rotation, the pseudo-modes retain toroidal displacements and can capture gravity/inertial modes in the inertial frequency regime $\omega<2\Omega_{\rm S}$ where the Coriolis and buoyancy forces are comparable.

The genuine normal modes of the planet can be expressed as superpositions of pseudo-modes, with the Coriolis force, centrifugal force, and oblateness acting to couple pseudo-modes of different $l$ and also within the same $l$. We employ a perturbation theory to solve for the normal modes\cite{Dahlen1998}, accounting for the mode self-coupling and mode-mode coupling induced by Saturn's rapid rotation. Angular selection rules mandate that $l=l_\alpha$ pseudo-modes can interact with pseudo-modes having $l_\beta=l_\alpha$ and $l_\beta=l_\alpha\pm2$. Because rotation preserves azimuthal symmetry, modes with distinct $m$ remain uncoupled.

Strictly speaking, the coupling implies that each normal mode receives contributions from an infinite chain of $l$ values. A given $m=-2$ normal mode for example requires knowledge of $m=-2$ pseudo-modes with $l=2,\,4,\,6,\,\ldots$ and a range of radial orders $n$. In practice, only a small number of the pseudo-mode coupling coefficients are significant and so the eigenvalue problem can be solved numerically after suitably truncating the matrix of coupling coefficients.
For computational expediency, our large samples of models presented in the main text (Figs.~\ref{fig.cavity_width}-\ref{fig.core_proxy}) only retain pseudo-modes of $l=2$, and only retain those with radial order $n=1-3$, where $n$ counts the number of nodes in the poloidal horizontal displacement that fall within the g~mode cavity (Extended Data Fig.~\ref{fig.eigs_best}-\ref{fig.eigs_cavity_width}). This excludes acoustic overtone (p) modes at high frequency as well as higher order g~modes and inertial modes at lower frequency $(n\gg1)$.

The resulting $m=-2$ pattern in our principal calculations (appearing all figures but  is a sequence of four modes which we label by the dominant pseudo-mode component of each normal mode using the scheme $_{\,\,\,l}^ms_n$, with $s$ a placeholder for the mode type g, f, or p. In order of decreasing frequency the relevant $m=-2$ modes are identified with \ensuremath{_{\,\,\,\,2}^{-2}g_1}\xspace, \ensuremath{_{\,\,\,\,2}^{-2}f}\xspace, \ensuremath{_{\,\,\,\,2}^{-2}g_2}\xspace, and \ensuremath{_{\,\,\,\,2}^{-2}g_3}\xspace. \ensuremath{_{\,\,\,\,2}^{-2}f}\xspace is identified as the pseudo f mode because it has the largest surface gravity perturbation. Even so, this pseudo-mode has hybrid character: its displacement eigenfunctions initially evanesce from the planetary surface like an f mode, but become oscillatory and attain large horizontal displacements in the deeper g~mode cavity.
Although the lowest frequency mode \ensuremath{_{\,\,\,\,2}^{-2}g_3}\xspace would resonate outside the C ring for acceptable models (Figs.~\ref{fig.cavity_width} and \ref{fig.ompat_j2n}), we include it to ensure that the frequencies of the three modes being compared to data vary below the level of 0.1\% as a function of the chosen truncation limit in $n$.

While the truncation choice $l=2$, $n\leq3$ is rather aggressive, we find that it has an insignificant effect for our ability to deduce gross structure.
Auditing $\sim\!2,000$ individual interior models with fully coupled calculations that include all $l\leq14$ and $n\leq10$, we find that the more aggressive truncation choice $l=2$, $n\leq3$ affects the frequencies of \ensuremath{_{\,\,\,\,2}^{-2}g_1}\xspace, \ensuremath{_{\,\,\,\,2}^{-2}f}\xspace, and \ensuremath{_{\,\,\,\,2}^{-2}g_2}\xspace by $0.1-0.3\%$. This frequency error is overwhelmed by the systematic frequency uncertainty introduced by the perturbation theory we use. The perturbation theory is accurate to $\mathcal O\,(\Omega_{\rm S}/\omega_{\rm dyn})^2$ so that the systematic uncertainty in frequencies in Saturn's rotating frame is of order $(\Omega_{\rm S}/\omega_{\rm dyn})^3\approx5\%$. It is this dominant systematic uncertainty that we apply in the likelihood calculation described in the following section.

Restricting the calculation to pseudo-modes with $l=2$ artificially excludes the possibility that one or more of the $l=2$ pseudo-modes can undergo an avoided crossing with a pseudo-mode of higher $l$, endowing that otherwise undetectable high-$l$ mode with a detectable gravity perturbation.
This process can generate an observable resonance where there would not have been one otherwise, and was invoked \cite{2014Icar..242..283F} to explain the trio of finely split ($\sim0.3\%$ in frequency) $m=-3$ density waves with mixed success. These avoided crossings are extremely sensitive to details of the interior model. Additional observable $m=-2$ modes do appear in $\approx20\%$ of our models when they are subjected to the fully coupled calculation. The process can be seen in Extended Data Fig.~\ref{fig.m2_many_l} where resonant coupling across $l$ creates columns of attracted $l>2$ modes at frequencies similar to \ensuremath{_{\,\,\,\,2}^{-2}g_2}\xspace or \ensuremath{_{\,\,\,\,2}^{-2}f}\xspace whose amplitudes are enhanced by the avoided crossings with these $l=2$ modes.
The fine ($\sim0.5\%$) frequency splitting between the Maxwell ringlet wave and W87.19 is therefore not an uncommon feature in the model spectra (vertical column at $\Omega_{\rm p}\approx1720\ {\rm deg\ d}^{-1}$), although the models systematically underestimate these two resonance frequencies in an absolute sense.
We note that observable fine splitting is equally common in the vicinity of \ensuremath{_{\,\,\,\,2}^{-2}f}\xspace (vertical column at $\Omega_{\rm p}\approx1840\ {\rm deg\ d}^{-1}$), although only a single wave W84.64 has been observed there.
A similar magnitude of frequency splitting is observed between the three $m=-3$ waves,
and Saturn's deep latitudinal differential rotation\cite{2019Sci...364.2965I,2019GeoRL..46..616G} may be a necessary ingredient to fully reproduce all of these instances of strong coupling\cite{2014Icar..242..283F}.
{The contribution that realistic differential rotation profiles make to the isolated mode frequencies that drive our main results are estimated in the Supplementary Information.}

\noindent{\bf Parameter estimation.}
To each interior model we assign a multivariate normal likelihood $\mathcal L=\mathcal L_{\rm grav}\mathcal L_{\rm seis}$ where
\begin{equation}
  \mathcal \ln \mathcal L_{\rm grav}=-\sum_{n=1,2}\delta J_{2n}^{-2}(J_{2n}^{\rm model}-J_{2n})^2
\end{equation}
and
\begin{equation}
  \mathcal \ln \mathcal L_{\rm seis}=-\sum_{i=1}^3\delta\Omega_{\rm p}^{-2}(\Omega_{{\rm p},i}^{\rm model}-\Omega_{{\rm p},i})^2.
\end{equation}
The gravity harmonics $J_{2n}$ are those derived from Cassini Grand Finale orbits\cite{2019Sci...364.2965I}, scaled to our assumed equatorial radius $60,268~{\rm km}$.
{From these we subtract the dynamical contributions estimated by Galanti \& Kaspi\cite{2021MNRAS.501.2352G} (their Table 1) to yield rigid-body components that can be meaningfully compared to our rigidly rotating models.}
{Although $J_2$ and $J_4$ are each measured to an absolute precision of order $10^{-8}$, and $J_6$ to order $10^{-7}$, the modeled harmonics $J_{2n}^{\rm model}$ from ToF suffer larger systematic shifts relative to the more accurate (but prohibitively slow for our purposes) Concentric Maclaurin Spheroids method\cite{2013ApJ...768...43H,2017A&A...606A.139N}.
Consequently we take our modeled harmonics as uncertain at the level of these dominant systematic offsets quantified by Nettelmann\cite{2017A&A...606A.139N} and adopt $\delta J_2=0.3~{\rm ppm}$, $\delta J_4=0.4~{\rm ppm}$, and $\delta J_6=0.7~{\rm ppm}$.
}

The $\Omega_{{\rm p},i}$ are observed pattern speeds of $m=-2$ spiral density waves.
Modeled pattern speeds $\Omega_{{\rm p},i}^{\rm model}$ are straightforward to compare once each observed wave has been identified with a given normal mode.
As described above, our acceptable models generally predict three $m=-2$ normal modes whose outer Lindblad resonances fall in the vicinity of the C ring (Fig.~\ref{fig.cavity_width}).
These three modes furthermore each have sufficient amplitude to generate a detectable density wave (Extended Data Fig.~\ref{fig.m2_many_l}).
We then make mode identifications based on frequency, identifying the interface/g~mode \ensuremath{_{\,\,\,\,2}^{-2}g_1}\xspace with W76.44\cite{2019Icar..319..599F}, the f-dominated mode \ensuremath{_{\,\,\,\,2}^{-2}f}\xspace with W84.64\cite{2019AJ....157...18H}, and the g~mode \ensuremath{_{\,\,\,\,2}^{-2}g_2}\xspace with either W87.19\cite{2014MNRAS.444.1369H} or the Maxwell ringlet wave\cite{2016Icar..279...62F}, whichever is closer to the predicted pattern speed.
The latter distinction is unimportant for our purposes because these two waves have indistinguishable pattern speeds relative to the 5\% accuracy achieved by our second-order perturbation theory. For $m=-2$ this 5\% systematic error corresponds to a pattern speed uncertainty $\delta\Omega_{\rm p}\sim90\ {\rm deg\ d}^{-1}$.

Equipped with the likelihood $\mathcal L$ and appropriate priors chosen for the free model parameters {\ensuremath{Z_{\rm out}}\xspace, \ensuremath{r_{\rm out}}\xspace, \ensuremath{Y^\prime_{\rm in}}\xspace and \ensuremath{f_{\rm ice}}\xspace}, we estimate the posterior probability density using emcee\cite{2013PSP..125..306F} v2.2.1. We assign uniform prior probability to {\ensuremath{Z_{\rm out}}\xspace over the range $[0,0.5]$, to \ensuremath{r_{\rm out}}\xspace over the range $[10^{-3}, 1]$, to \ensuremath{Y^\prime_{\rm in}}\xspace over the range $[0.275, 1]$, and to \ensuremath{f_{\rm ice}}\xspace over $[0, 1]$}.
Samples varying $T_1$ impose a Gaussian prior probability with mean {135 K and standard deviation 5 K}.
We also compute a case in which Saturn's assumed rotation rate is allowed to vary, imposing a Gaussian prior on the smallness parameter $m_{\rm rot}=(\Omega_{\rm S}/\omega_{\rm dyn})^2$ with mean $0.14218$ and standard deviation $0.00071$, corresponding to a virtually Gaussian period distribution with standard deviation 1m 35s about the mean period 10h 33m 38s.

Extended Data Table~\ref{tab.models} summarizes and compares all samples, reporting estimates of parameters and derived quantities in each case.
We also tabulate the maximum posterior probability $\hat P$ realized in each sample along with the individual gravity and seismology likelihoods $\hat{\mathcal L}_{\rm grav}$ and $\hat{\mathcal L}_{\rm seis}$ of that best model.
Denoting by $N_{\rm par}$ the number of free parameters in each model, the Akaike information criterion\cite{1100705} ${\rm AIC}=2N_{\rm par}-2\ln\hat{\mathcal L}$ is minimized for the {4-parameter} linear and sigmoid cases, indicating that only the 4 parameters {\ensuremath{Z_{\rm out}}\xspace, \ensuremath{r_{\rm out}}\xspace, \ensuremath{Y^\prime_{\rm in}}\xspace and \ensuremath{f_{\rm ice}}\xspace} controlling the heavy element distribution are essential to reproduce the data.
{Of particular note is that samples allowing \ensuremath{r_{\rm in}}\xspace to vary freely yield no improvement to the fit, and so there is no evidence in particular for a chemically homogeneous inner core region.
The possibility of superadiabatic thermal stratification in the deep gradient region is also unlikely to change our findings (see Supplementary Information).
Of the favored four-parameter models, we adopt the linear $Z,\ensuremath{Y^\prime}\xspace$ case as our baseline because it minimizes the AIC and furthermore produces the best fitting individual models from the perspective of both the gravity and seismology constraints.
We note however that, with a relative likelihood of 0.81, the sigmoid $Z,\ensuremath{Y^\prime}\xspace$ case performs only marginally worse.
}


\section*{Additional Information}
\begin{itemize}
 \item {\bf Acknowledgments} C.M. thanks David Stevenson for comments and acknowledges support from the Division of Geological and Planetary Sciences at Caltech. J.F. is thankful for support through an Innovator Grant from The Rose Hills Foundation, and the Sloan Foundation through grant FG-2018-10515.
 \item {\bf Author Contributions} C.M. developed the planetary models, performed the calculations and analysis, and led the preparation of the manuscript. J.F. developed the original oscillation code, contributed to the interpretation of the results, and helped to write the manuscript.
 \item {\bf Competing Interests} The authors declare that they have no
competing financial interests.
 \item {\bf Correspondence} Correspondence and requests for materials
should be addressed to C.M.~(email: chkvch@caltech.edu).
\item Supplementary Information is available for this paper.
\end{itemize}

 {\bf Data Availability. } A representative subset of the interior models generated in the course of this work are available upon request.

 {\bf Code Availability. } The planetary structure and theory of figures code used to create the planetary models is available at \verb+https://github.com/chkvch/alice+. The oscillation code and ancillary code related to the analysis are available upon request.

\section{Extended Data}

\setcounter{figure}{0}
\setcounter{table}{0}

\captionsetup[figure]{name=Extended Data Figure}
\captionsetup[table]{name=Extended Data Table}

\begin{table}[htbp]
  \scriptsize
  \begin{tabular}{lccccccc}
  & \multicolumn{7}{c}{Model comparison} \\
  \cline{2-8}
  Sample                        & $N_{\rm par}$       & $\ln\hat P$         & $\ln\hat{\mathcal L}_{\rm grav}$        & $\ln\hat{\mathcal L}_{\rm seis}$        & AIC                 & Relative likelihood & \\
  \hline
  \vspace{1mm}
  {\bf Linear }${\pmb{Z,Y^\prime}}$& 4                   & -0.19               & -2.4e-03            & -1.8e-01            & 8.37                & {\bf 1.00          }\!---Baseline\\
  \quad Vary $r_{\rm in}$       & 5                   & -0.19               & -1.1e-02            & -1.8e-01            & 10.38               & 0.37                \\
  \quad Vary $T_1$              & 5                   & -0.19               & -2.2e-03            & -1.8e-01            & 10.37               & 0.37                \\
  \quad Vary rotation           & 5                   & -0.18               & -5.7e-03            & -1.8e-01            & 10.37               & 0.37                \\
  \quad Superadiabatic          & 5                   & -0.19               & -7.8e-03            & -1.8e-01            & 10.39               & 0.37                \\
  \vspace{1mm}
  \quad Neglect $J_{2n}$        & 4                   & -0.15               & -2.4e+04            & -1.5e-01            & n/a                 & n/a                 \\
  \vspace{1mm}
  {\bf Sigmoid} $\pmb{Z,Y^\prime}$& 4                   & -0.40               & -1.3e-02            & -3.9e-01            & 8.80                & {\bf 0.81           }\\
  \quad Vary $r_{\rm in}$       & 5                   & -0.40               & -1.1e-02            & -3.9e-01            & 10.80               & 0.30                \\
  \quad Vary $T_1$              & 5                   & -0.31               & -3.8e-02            & -2.7e-01            & 10.63               & 0.32                \\
  \quad Vary rotation           & 5                   & -0.36               & -1.7e-02            & -3.4e-01            & 10.72               & 0.31                \\
  \quad Superadiabatic          & 5                   & -0.39               & -9.2e-02            & -2.9e-01            & 10.77               & 0.30                \\
  \vspace{1mm}
  \quad Neglect $J_{2n}$        & 4                   & -0.18               & -9.0e+04            & -1.8e-01            & n/a                 & n/a                 \\
  \end{tabular}\vspace{5mm}\\\\
  \begin{tabular}{lccccccccc}
  & \multicolumn{9}{c}{Parameters} \\
  \cline{2-10}
  Sample                        & $r_{\rm in}\ (R_{\rm S})$     & $r_{\rm out}\ (R_{\rm S})$    & $Z_1$                         & $Y_2^\prime$                  & $f_{\rm ice}$                 & $T_1\ (\rm K)$                & $m_{\rm rot}$                 & $\nabla-\nabla_{\rm ad}$      & \\
  \hline
  \vspace{1mm}
  {\bf Linear }${\pmb{Z,Y^\prime}}$& 0                             & $0.59\pm0.01$                 & $0.041\pm0.009$               & $0.91{\rm -}1.00$             & $0.13{\rm -}0.96$             & 135                           & 0.1422                        & 0                             & \\
  \quad Vary $r_{\rm in}$       & $0.01{\rm -}0.18$             & $0.59\pm0.02$                 & $0.046\pm0.018$               & $0.68{\rm -}0.98$             & $0.06{\rm -}0.94$             & ---                           & ---                           & ---                           & \\
  \quad Vary $T_1$              & ---                           & $0.60\pm0.01$                 & $0.036\pm0.012$               & $0.82{\rm -}0.99$             & $0.07{\rm -}0.93$             & $130.5{\rm -}136.3$           & ---                           & ---                           & \\
  \quad Vary rotation           & ---                           & $0.60\pm0.01$                 & $0.036\pm0.012$               & $0.81{\rm -}0.99$             & $0.06{\rm -}0.92$             & ---                           & $0.1424\pm0.0006$             & ---                           & \\
  \quad Superadiabatic          & ---                           & $0.59\pm0.01$                 & $0.039\pm0.007$               & $0.92{\rm -}1.00$             & $0.20{\rm -}0.90$             & ---                           & ---                           & $0.01{\rm -}0.48$             & \\
  \vspace{1mm}
  \quad Neglect $J_{2n}$        & ---                           & $0.42{\rm -}0.68$             & $0.005{\rm -}0.107$           & $0.56{\rm -}0.99$             & $0.04{\rm -}0.85$             & ---                           & ---                           & ---                           & \\
  \vspace{1mm}
  {\bf Sigmoid} $\pmb{Z,Y^\prime}$& ---                           & $0.70\pm0.02$                 & $0.057\pm0.015$               & $0.81{\rm -}0.93$             & $0.05{\rm -}0.94$             & ---                           & ---                           & ---                           & \\
  \quad Vary $r_{\rm in}$       & $0.00{\rm -}0.16$             & $0.67\pm0.03$                 & $0.044\pm0.016$               & $0.64{\rm -}0.89$             & $0.05{\rm -}0.95$             & ---                           & ---                           & ---                           & \\
  \quad Vary $T_1$              & ---                           & $0.70\pm0.02$                 & $0.058\pm0.025$               & $0.74{\rm -}0.98$             & $0.06{\rm -}0.94$             & $130.9{\rm -}140.1$           & ---                           & ---                           & \\
  \quad Vary rotation           & ---                           & $0.69\pm0.02$                 & $0.047\pm0.020$               & $0.73{\rm -}0.93$             & $0.06{\rm -}0.95$             & ---                           & $0.1422\pm0.0007$             & ---                           & \\
  \quad Superadiabatic          & ---                           & $0.71\pm0.03$                 & $0.065\pm0.019$               & $0.84{\rm -}0.99$             & $0.06{\rm -}0.93$             & ---                           & ---                           & $0.02{\rm -}0.43$             & \\
  \vspace{1mm}
  \quad Neglect $J_{2n}$        & ---                           & $0.53{\rm -}0.77$             & $0.004{\rm -}0.109$           & $0.34{\rm -}0.97$             & $0.06{\rm -}0.94$             & ---                           & ---                           & ---                           & \\
  \end{tabular}\vspace{5mm}\\\\
  \begin{tabular}{lccccccccc}
  & \multicolumn{9}{c}{Derived quantities} \\
  \cline{2-10}
  Sample                        & $M_{0.5}\ (M_{\rm E})$                        & $R_{0.5}\ (R_{\rm S})$                        & $M_{Z,\rm tot}$                        & $N_{\rm max}/\omega_{\rm dyn}$                         & $Z_2$                         & $Y_1^\prime$                    & $M_{\rm stab}$                        & $M_{Z,\rm stab}$                       & \\
  \hline
  \vspace{1mm}
  {\bf Linear }${\pmb{Z,Y^\prime}}$& $18.6\pm1.5$                  & $0.32\pm0.01$                 & $19.1\pm1.0$                  & $1.97\pm0.05$                 & $0.64{\rm -}0.98$             & $0.16\pm0.02$                 & $55.1\pm1.7$                  & $17.4\pm1.2$                  & \\
  \quad Vary $r_{\rm in}$       & $19.7\pm3.2$                  & $0.30{\rm -}0.39$             & $18.7\pm1.1$                  & $1.96\pm0.09$                 & $0.65\pm0.12$                 & $0.07{\rm -}0.19$             & $55.9\pm3.2$                  & $16.9\pm1.4$                  & \\
  \quad Vary $T_1$              & $18.5\pm1.7$                  & $0.32\pm0.01$                 & $18.8\pm1.0$                  & $1.94\pm0.05$                 & $0.63{\rm -}0.97$             & $0.16\pm0.02$                 & $56.2\pm2.0$                  & $17.4\pm1.2$                  & \\
  \quad Vary rotation           & $18.6\pm1.7$                  & $0.32\pm0.01$                 & $18.8\pm1.0$                  & $1.94\pm0.05$                 & $0.63{\rm -}0.97$             & $0.16\pm0.02$                 & $56.4\pm2.1$                  & $17.5\pm1.2$                  & \\
  \quad Superadiabatic          & $18.2\pm1.2$                  & $0.32\pm0.01$                 & $19.9\pm1.0$                  & $1.98\pm0.05$                 & $0.70{\rm -}0.99$             & $0.16\pm0.01$                 & $55.0\pm1.4$                  & $18.3\pm1.2$                  & \\
  \vspace{1mm}
  \quad Neglect $J_{2n}$        & $15.2{\rm -}23.7$             & $0.31{\rm -}0.39$             & $18.1\pm1.4$                  & $1.92\pm0.07$                 & $0.64{\rm -}0.98$             & $0.08{\rm -}0.25$             & $24.0{\rm -}69.4$             & $8.4{\rm -}20.0$              & \\
  \vspace{1mm}
  {\bf Sigmoid} $\pmb{Z,Y^\prime}$& $21.0\pm2.9$                  & $0.34\pm0.02$                 & $18.4\pm1.0$                  & $2.05\pm0.06$                 & $0.60\pm0.10$                 & $0.12\pm0.03$                 & $69.8\pm2.9$                  & $17.0\pm1.1$                  & \\
  \quad Vary $r_{\rm in}$       & $15.8{\rm -}24.7$             & $0.30{\rm -}0.37$             & $17.1{\rm -}20.5$             & $2.14\pm0.11$                 & $0.62\pm0.09$                 & $0.09{\rm -}0.19$             & $58.3{\rm -}72.9$             & $15.6{\rm -}19.5$             & \\
  \quad Vary $T_1$              & $21.4\pm3.9$                  & $0.34\pm0.03$                 & $18.5\pm1.0$                  & $2.06\pm0.06$                 & $0.59\pm0.12$                 & $0.12\pm0.04$                 & $69.7\pm2.7$                  & $17.0\pm1.3$                  & \\
  \quad Vary rotation           & $19.7\pm3.0$                  & $0.33\pm0.02$                 & $18.8\pm1.2$                  & $2.07\pm0.06$                 & $0.65\pm0.12$                 & $0.14\pm0.03$                 & $69.3\pm2.6$                  & $17.6\pm1.4$                  & \\
  \quad Superadiabatic          & $21.9\pm3.3$                  & $0.35\pm0.03$                 & $17.3{\rm -}24.0$             & $2.09\pm0.06$                 & $0.62\pm0.13$                 & $0.11\pm0.04$                 & $71.5\pm3.4$                  & $18.6\pm2.2$                  & \\
  \vspace{1mm}
  \quad Neglect $J_{2n}$        & $13.6{\rm -}27.7$             & $0.29{\rm -}0.41$             & $18.5\pm1.4$                  & $2.05\pm0.06$                 & $0.43{\rm -}0.96$             & $0.05{\rm -}0.27$             & $39.2{\rm -}79.7$             & $11.3{\rm -}20.5$             & \\
  \end{tabular}
  \caption{\label{tab.models}
  {\bf Results of seismology/gravity retrievals for different parameterizations of Saturn's interior structure.}
  Physical properties are reported in terms of their means and standard deviations.
  In cases where distributions are significantly non-Gaussian, ranges corresponding to 5\% and 95\% quantiles are reported instead.
  Blank entries take the same values as in the baseline case.
  }
  \end{table}

\begin{figure}[p]
  \centering
  \includegraphics[width=\textwidth]{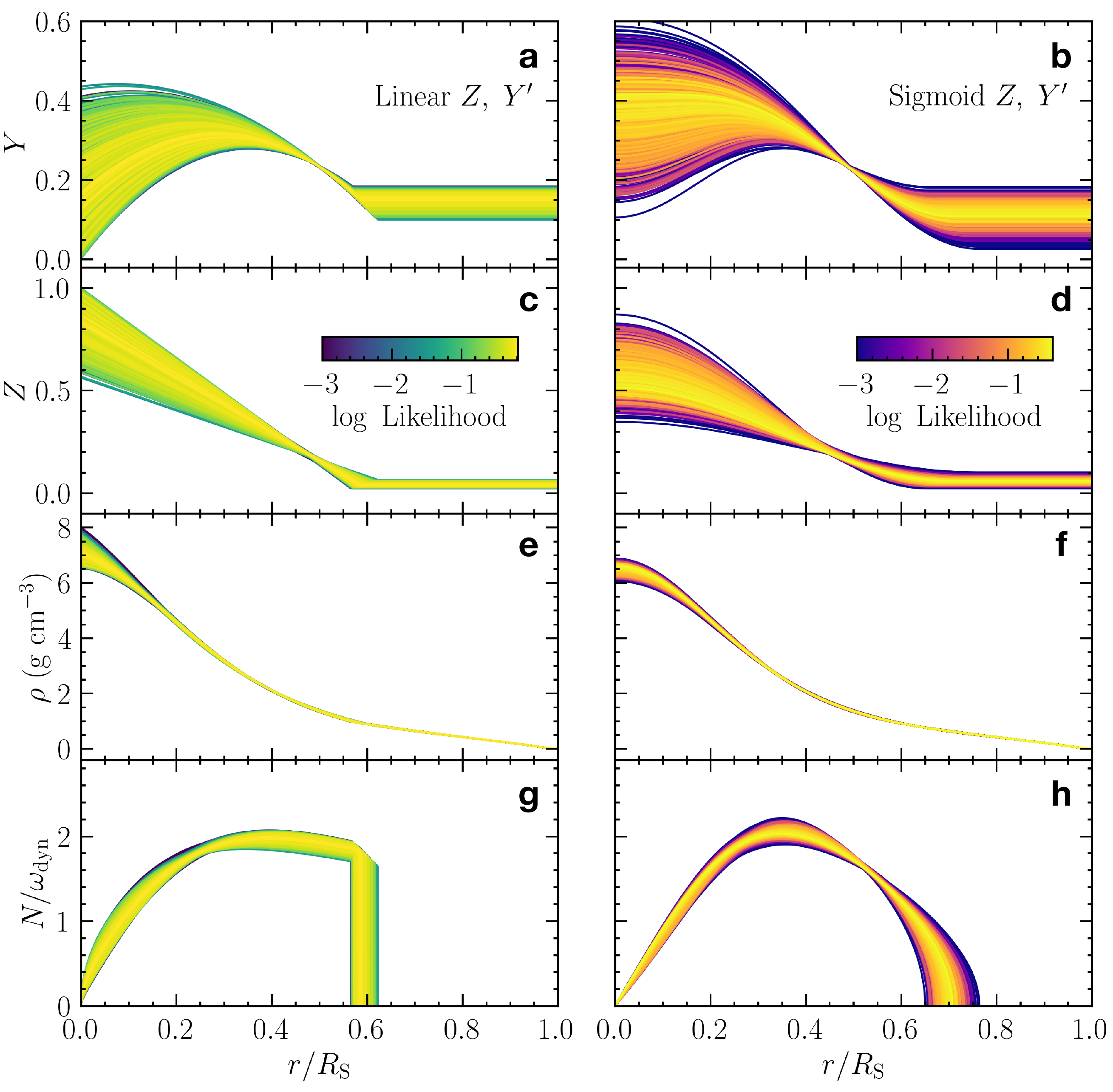}
  \caption{\label{fig.profiles_comparison}
    {\bf Comparison of assumed shapes for Saturn's composition gradient.}
    Helium distributions for our baseline linear composition profiles from Eq.~\ref{eq.z_profile_linear} ({\bf a}) are compared with those assuming sigmoid $Z(r)$ and $\ensuremath{Y^\prime}\xspace(r)$ in the transition region ({\bf b}).
    The corresponding profiles of the heavy element mass fraction ({\bf c}-{\bf d}), mass density ({\bf e}-{\bf f}), and Brunt-V\"ais\"al\"a frequency ({\bf g}-{\bf h}) are also shown.
    {These are 1,024 randomly selected models from each sample, colored by log likelihood.}
  }
\end{figure}

\begin{figure}[p]
  \centering
  \includegraphics[width=0.8\textwidth]{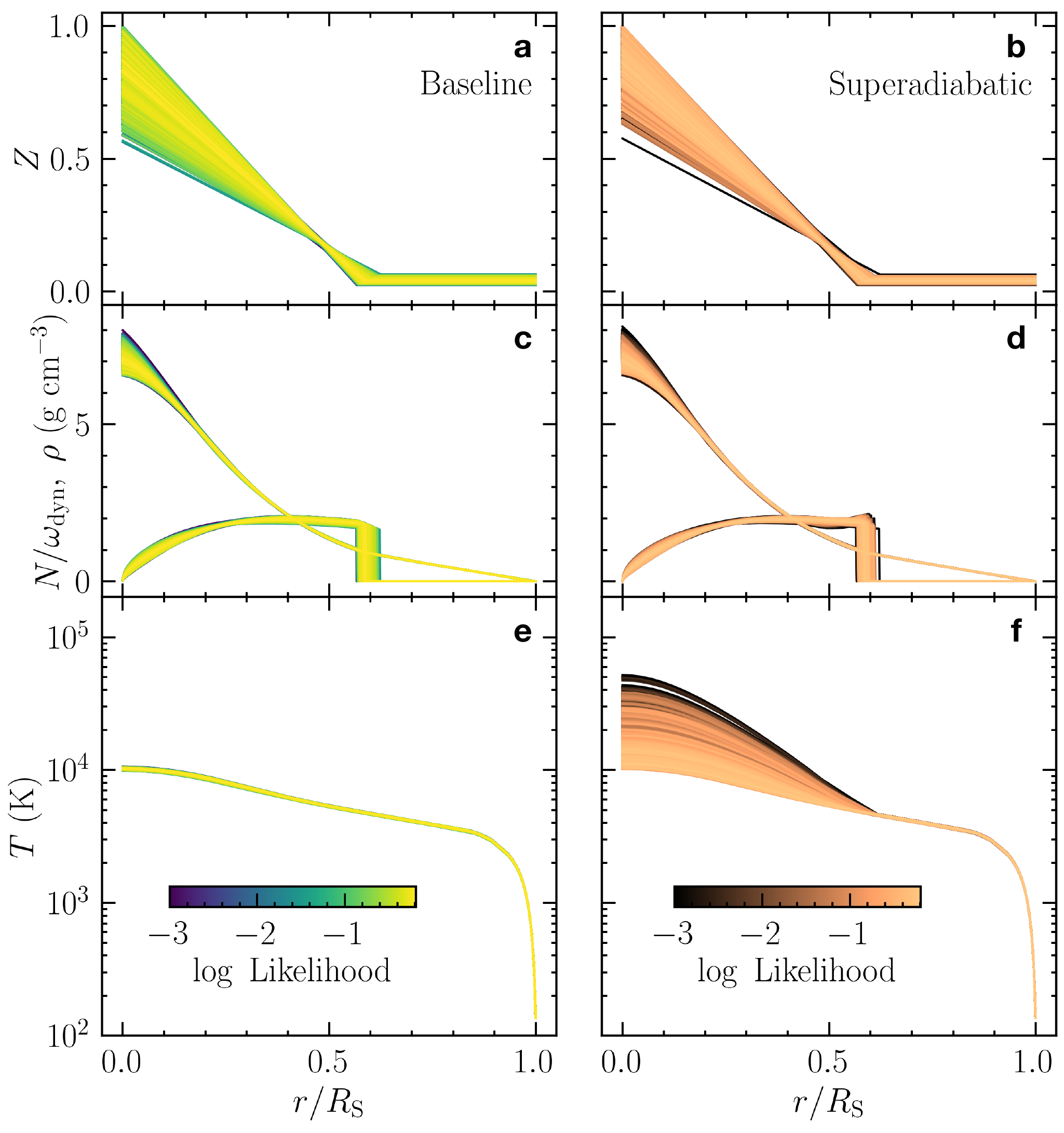}
  \caption{\label{fig.superad}
  {\bf Effect of superadiabatic thermal stratification.}
  Heavy element distributions for our baseline case ({\bf a}) are compared with those for our superadiabatic case ({\bf b}).
  ({\bf c}-{\bf d}) show profiles of Brunt-V\"ais\"al\"a frequency and mass density, and ({\bf e}-{\bf f}) show temperature profiles.
  {These are 1,024 randomly selected models from each sample, colored by log likelihood.}
  }
\end{figure}

\begin{figure}[p]
  \centering
  \includegraphics[width=0.7\textwidth]{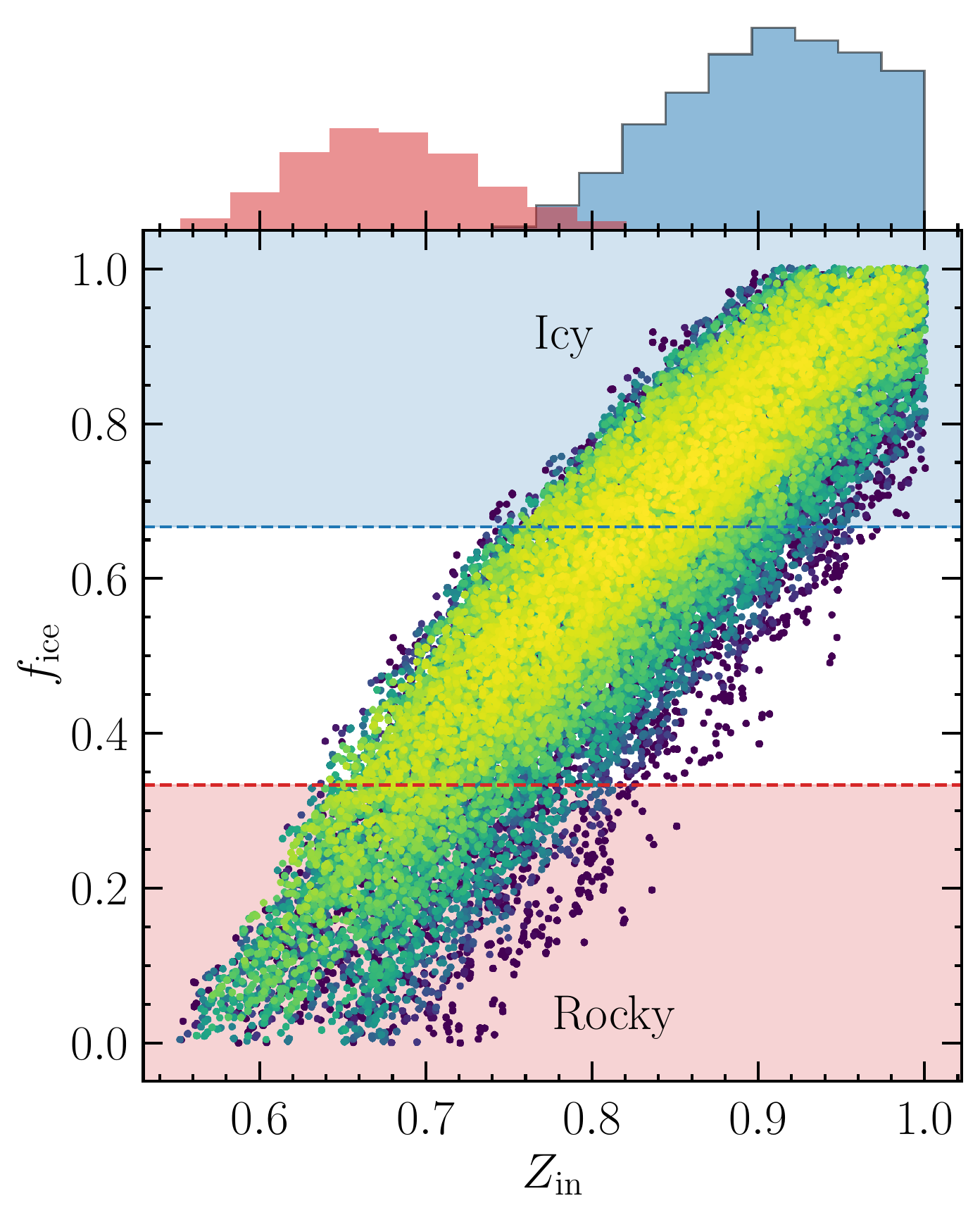}
  \caption{\label{fig.z2-fice}
  {
  {\bf Relationship between ice to rock mass fraction, $f_{\rm ice}$, and predicted central heavy element mass fraction (ice plus rock), \ensuremath{Z_{\rm in}}\xspace.}
  The red histogram shows the distribution of \ensuremath{Z_{\rm in}}\xspace in models with $\ensuremath{f_{\rm ice}}\xspace < 1/3$; the blue histogram shows the same for models with $\ensuremath{f_{\rm ice}}\xspace > 2/3$.
  Models are from the baseline case and colored by log likelihood as in Figs.~\ref{fig.profiles}-\ref{fig.core_proxy}. For models with predominantly icy cores, the preferred value of \ensuremath{Z_{\rm in}}\xspace is near unity, and vice versa.
  }
  }
\end{figure}

\begin{figure}[p]
  \centering
  \includegraphics[width=0.6\textwidth]{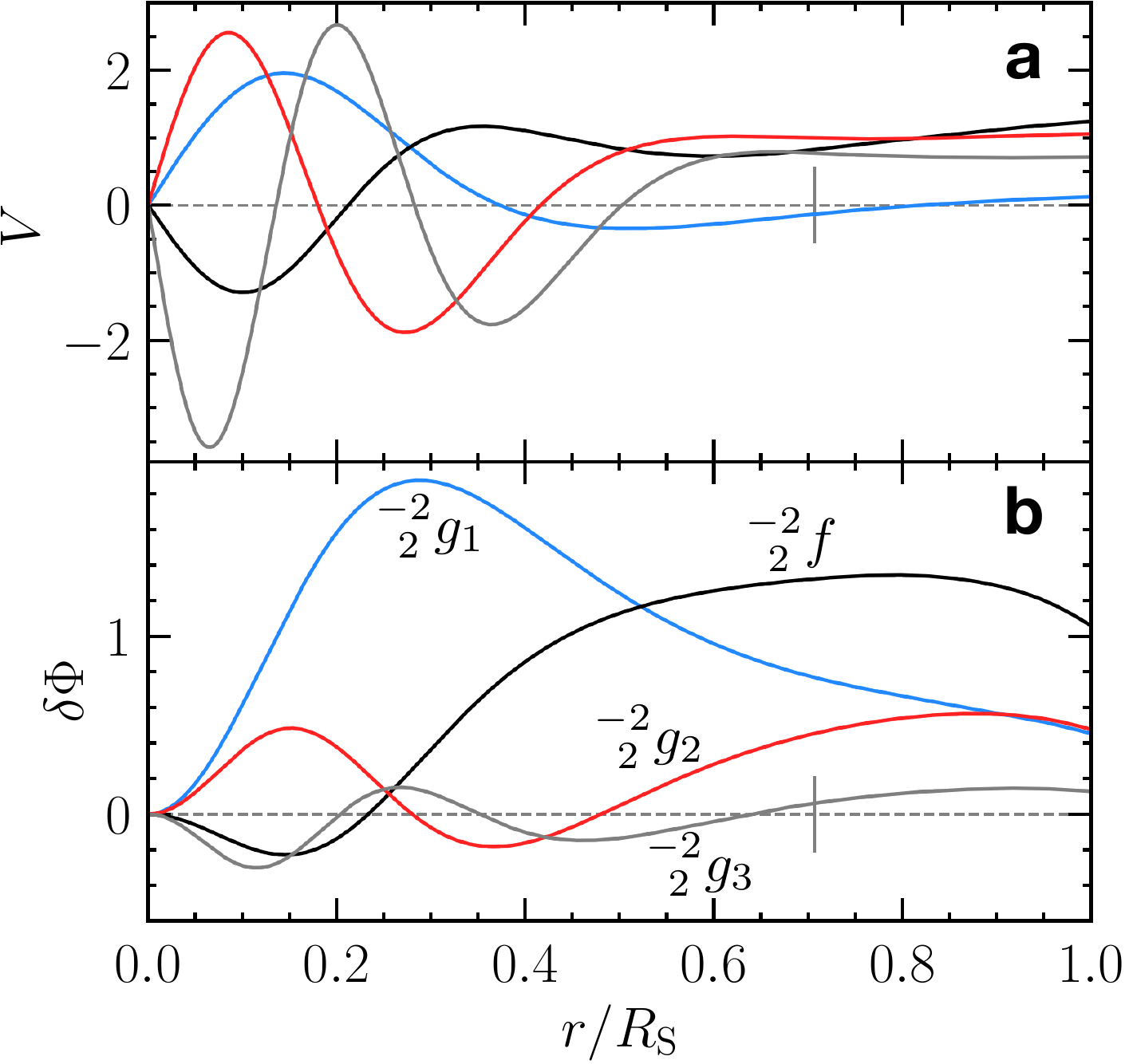}
  \caption{\label{fig.eigs_best}
  {\bf Eigenfunctions of $m=-2$, $l=2$ pseudo-modes in Saturn.}
  {\bf a,} Poloidal component of the horizontal displacement perturbation as a function of radius.
  {\bf b,} Gravitational potential perturbation as a function of radius.
  Vertical line segments mark the outer boundary of the g~mode cavity.
  }
\end{figure}

\begin{figure}[p]
  \centering
  \includegraphics[width=0.6\textwidth]{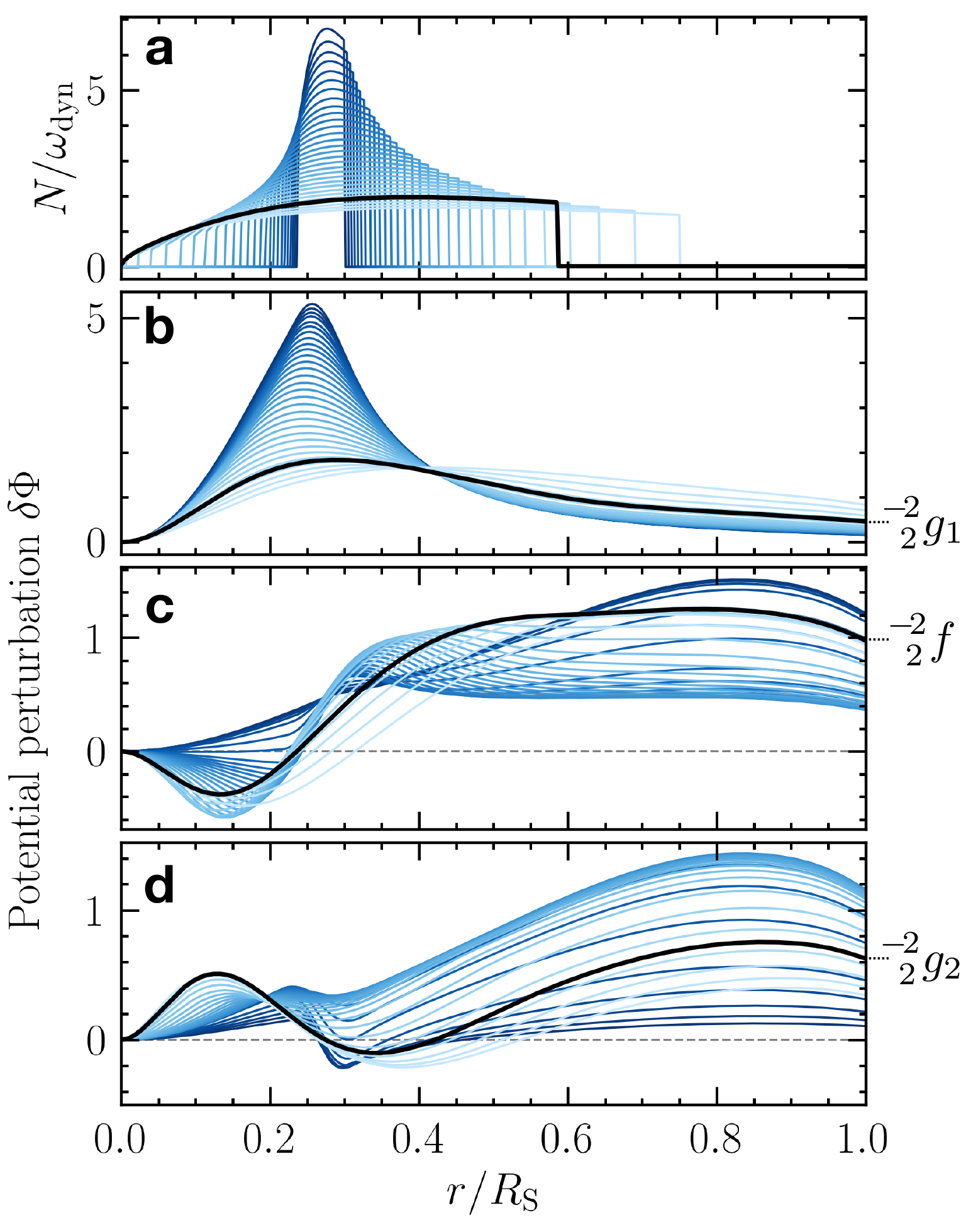}
  \caption{\label{fig.eigs_cavity_width}
  {\bf Eigenfunctions of $m=-2$, $l=2$ pseudo-modes as a function of g~mode cavity width.}
  This is the sequence of interior models from Fig.~\ref{fig.cavity_width}, with our most likely model from the sample of Figs.~\ref{fig.profiles}-\ref{fig.ompat_j2n} plotted in black.
  {\bf a,} Brunt-V\"ais\"al\"a frequency as a function of radius.
  {\bf b-d,} Gravitational potential perturbations associated {with the three highest frequency pseudo-modes in descending order.
  The identifications \ensuremath{_{\,\,\,\,2}^{-2}f}\xspace and \ensuremath{_{\,\,\,\,2}^{-2}g_2}\xspace hold for the best model (heavy black curves) but not necessarily others: moderate g~mode cavity widths bring the modes in ({\bf c}) and ({\bf d}) farther away from an avoided crossing, causing the mode in ({\bf d}) to become more f~mode-like and ({\bf c}) more g~mode-like.
  }
  }
\end{figure}

\begin{figure}[p]
  \centering
  \includegraphics[width=\textwidth]{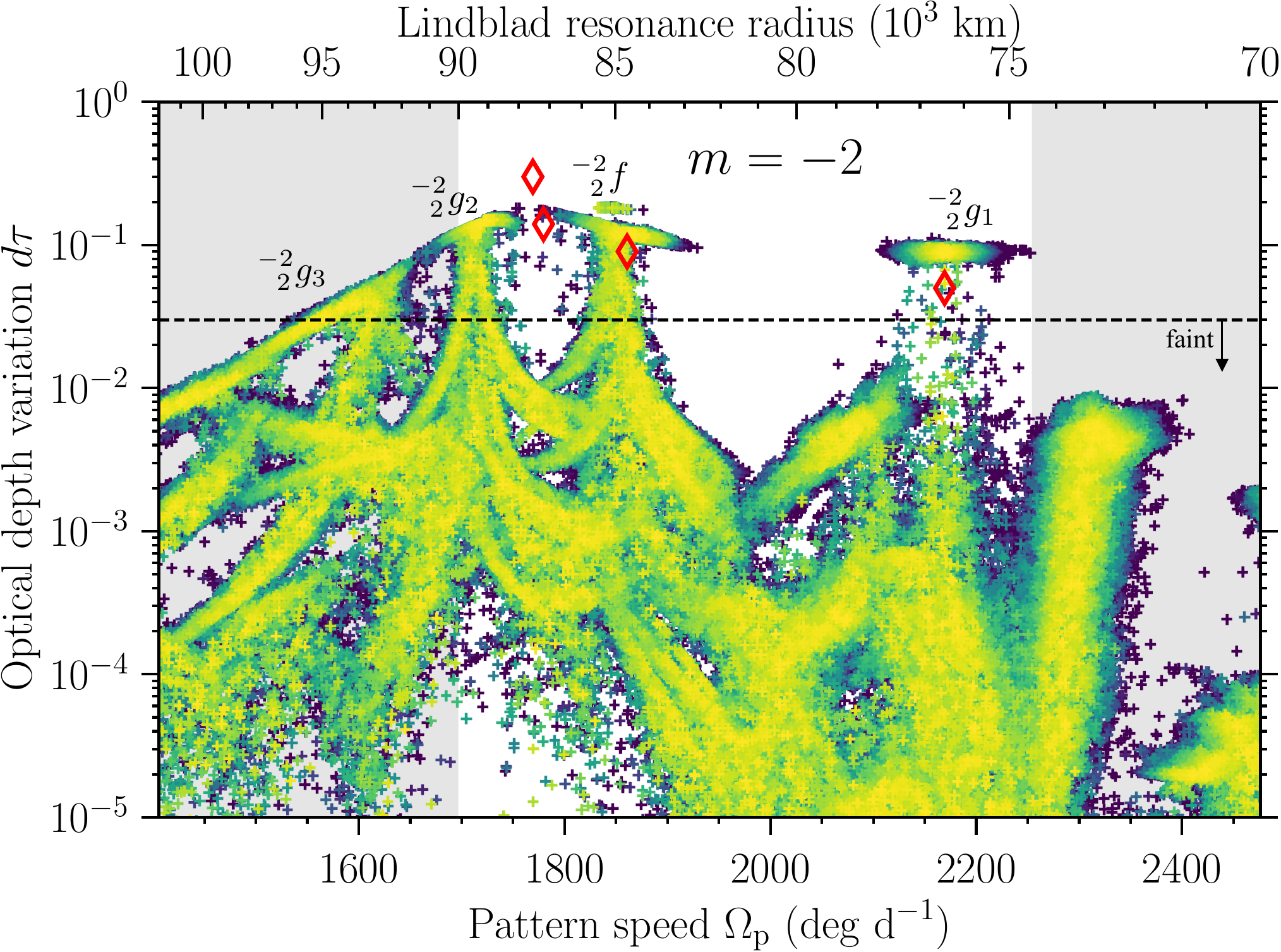}
  \caption{\label{fig.m2_many_l}
  {\bf Saturn's $m=-2$ mode spectrum including coupling across pseudo-modes of different $l$.}
  Predicted semi-amplitude $d\tau$ of optical depth variations near outer Lindblad resonances of density waves in Saturn's rings is plotted as a function of frequency $\Omega_{\rm p}$ and resonance radius in the ring plane.
  Colors indicate interior model likelihood with the same mapping as in Figs.~\ref{fig.profiles} and \ref{fig.core_proxy}.
  Red diamond symbols mark the frequencies and approximate amplitudes of spiral density waves observed at $m=-2$ outer Lindblad resonances.
  From left to right these are the Maxwell ringlet wave\cite{2016Icar..279...62F}, W87.19\cite{2014MNRAS.444.1369H}, W84.64\cite{2019AJ....157...18H}, and W76.44\cite{2019Icar..319..599F}.
  }
\end{figure}

\begin{figure}[p]
  \centering
  \includegraphics[width=0.75\textwidth]{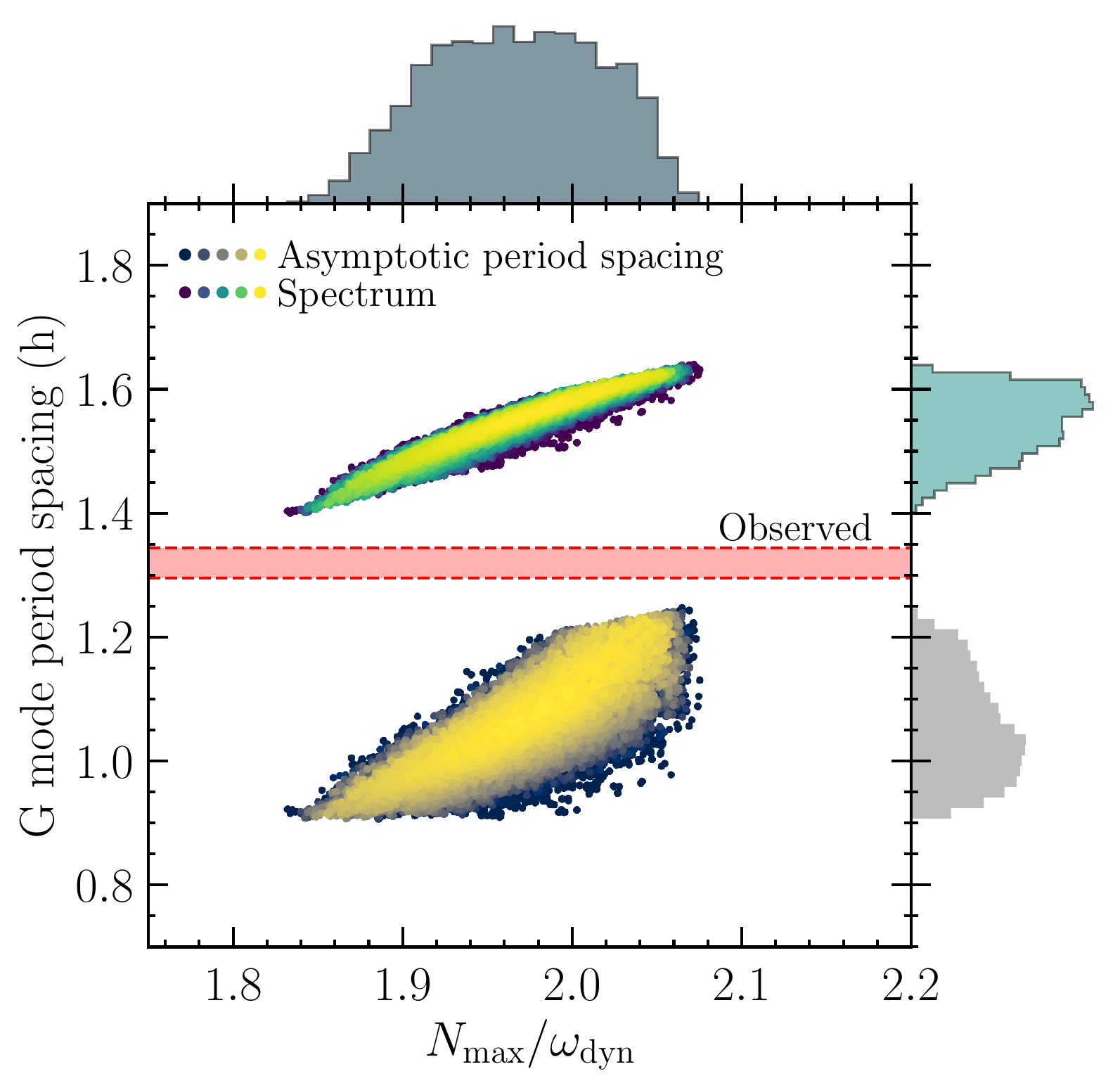}
  \caption{\label{fig.period_spacing}
  {\bf Estimating Saturn's g~mode period spacing.}
  The g~mode period spacing is plotted against $N_{\rm max}$ for each model in our {baseline} sample, with color mapped to log likelihood. The asymptotic spacing is given by Eq.~\ref{eq.brunt} in the Supplementary Information and the {exact} spacing labeled ``spectrum'' is calculated from the \ensuremath{_{\,\,\,\,2}^{-2}g_1}\xspace and \ensuremath{_{\,\,\,\,2}^{-2}g_2}\xspace mode frequencies.
  The lower (upper) bound for the observed period {spacing} assumes that W76.44 and W87.19 (Maxwell) are generated by $l=2$ Saturn g~modes of consecutive radial order.
  }
\end{figure}

\begin{figure}[p]
  \centering
  \includegraphics[width=0.752\textwidth]{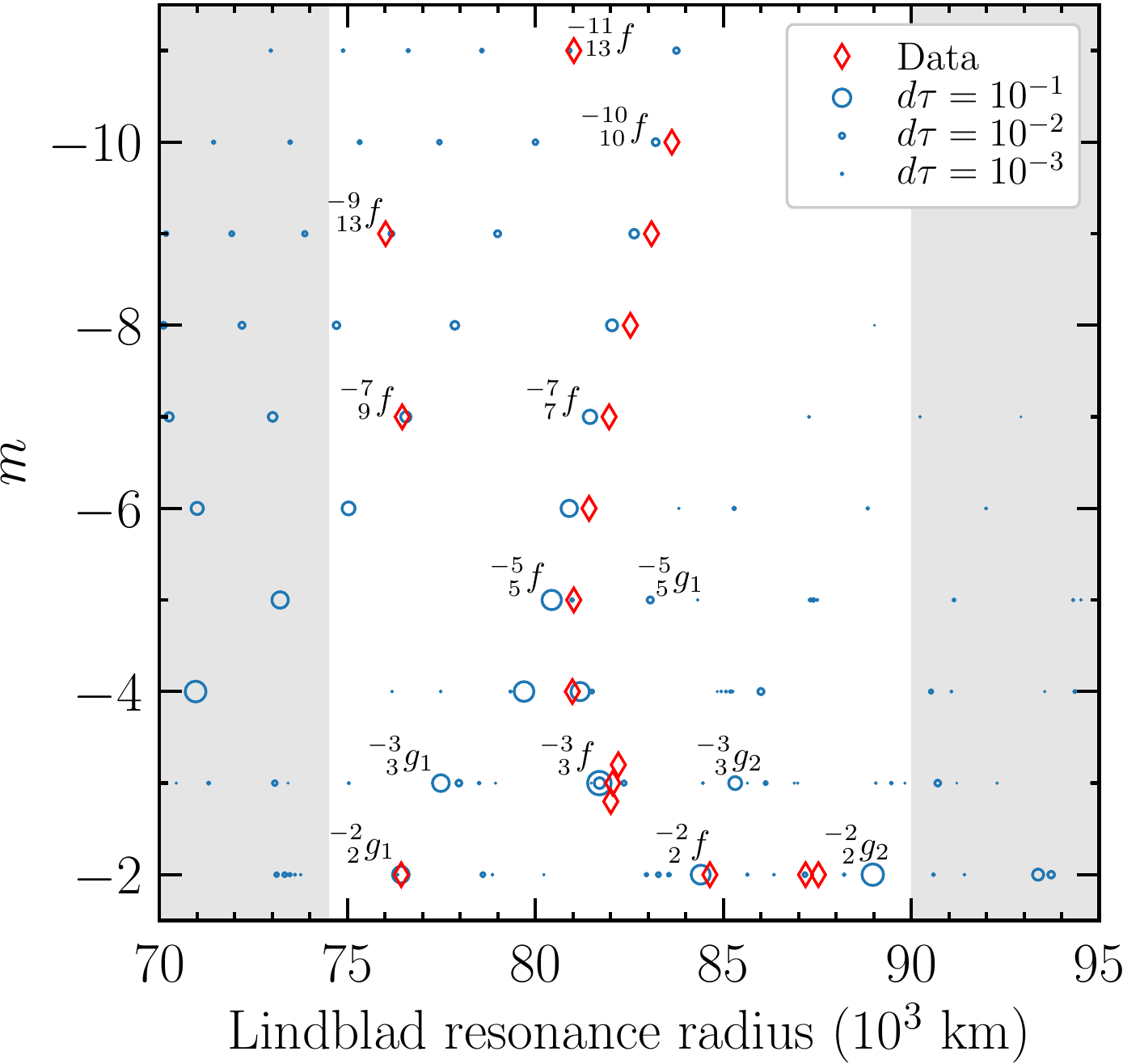}
  \caption{\label{fig.m_rl}
  {\bf Saturn's oscillation spectrum across $m$.}
  Locations of observed patterns at outer Lindblad resonances in the C ring (diamonds) are compared to modes of our preferred Saturn model (circles, with marker size mapping to estimated wave optical depth amplitude).
  Select modes are labeled by their dominant pseudo-mode component.
  }
\end{figure}

\begin{figure}[p]
  \centering
  \includegraphics[width=\textwidth]{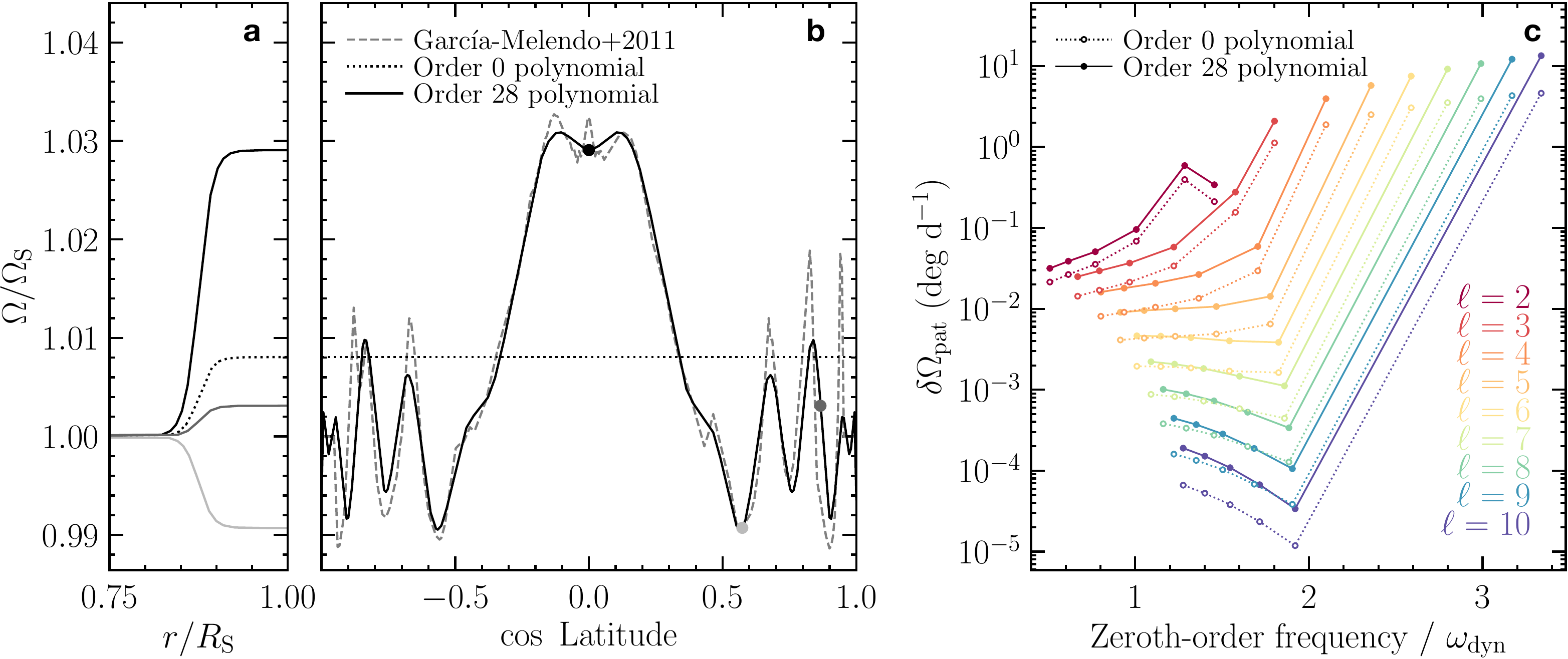}
  \caption{\label{fig.firstord_dr}
  {\bf Effect of deep zonal winds on sectoral mode frequencies.}
  {\bf a}, two rotation laws of the form~(\ref{eq.rotation_law}) as a function of spherical radius inside Saturn.
  {\bf b}, the same as a function of latitude at the cloud level.
  Observed winds are shown in dashed grey.
  For the nontrivial expansion labled ``Order 28 polynomial'' radial profiles ({\bf a}) are shown at three latitudes marked with filled circles in ({\bf b}).
  {\bf c}, first-order perturbations to $l=-m$ mode pattern speeds induced by the differential rotation.
  }
\end{figure}

\pagebreak
\section*{Supplementary Information}
\noindent{\bf The g~mode period spacing.}
Because the C ring reveals two g-dominated modes of consecutive radial order, we may estimate Saturn's g~mode period spacing, a fundamental diagnostic of internal stratification that has been applied to measure convective core sizes in main sequence stars\cite{2008AN....329..529M,2010Natur.464..259D} and red giants\cite{2011Natur.471..608B}.
In the asymptotic limit $n\gg1$, and ignoring the effect of rotation, g~modes are expected to be uniformly spaced in period by a separation\cite{1979nos..book.....U}
\begin{equation}
\label{eq.period_spacing}
  \Delta P_g^{\rm asy}=\frac{2\pi^2}{\sqrt{l(l+1)}}\left(\int_g N\,d\ln r\right)^{-1},
\end{equation}
where the integral is carried out over the g~mode cavity $N>0$.
Extended Data Fig.~\ref{fig.period_spacing} displays this asymptotic period spacing for models in our baseline sample, comparing it to the {exact} spacing between the \ensuremath{_{\,\,\,\,2}^{-2}g_1}\xspace and \ensuremath{_{\,\,\,\,2}^{-2}g_2}\xspace periods derived from the calculated mode spectra. The asymptotic spacing underestimates the {exact} spacing by typically 50\%. The imperfect correspondence between the two is unsurprising given that the modes in question are far from the asymptotic limit, are affected by rotation, and are both interacting strongly with the nearby f mode.

It is significant, however, that the models universally over-predict the period spacing compared to the spacing inferred directly from observations.
We calculate the latter as the period difference, in the planet's rotating frame, between W76.44 and W87.19 (or the nearby Maxwell ringlet wave). The inertial pattern speeds $\Omega_{\rm p}$ are related to the frequency $\omega_\alpha$ in the planet's frame by
\begin{equation}
  \omega_\alpha=-m\Omega_{\rm p,\alpha}+m\Omega_{\rm S}.
\end{equation}
Here we consider a broad range of bulk rotation rates $\Omega_{\rm S}$ corresponding to periods between 10h32m\cite{2015Natur.520..202H,2019ApJ...879...78M} and 10h39m\cite{1981GeoRL...8..253D}.
The ambiguity in identifying either W87.19 or Maxwell with the higher-order g~mode dominates the period spacing uncertainty, yielding a final estimate of {1.30 to 1.34 h} for the observed spacing.
Our numerical mode spectra on the other hand predict {1.44 to 1.61 h}, an overestimate also apparent in Fig.~\ref{fig.ompat_j2n} and Extended Data Fig.~\ref{fig.m2_many_l} where our models on average satisfy the pattern speed of W76.44 but under-predict those of W87.19/Maxwell.
The inability of our model to satisfy the observed spacing may indicate that a different functional from from the somewhat arbitrary parameterizations chosen in Eqs.~\ref{eq.z_profile_linear}-\ref{eq.z_profile_sigmoid} might be necessary to capture Saturn's deep composition profile.
{We note that the smooth models assuming sigmoid radius dependence in $Z(r)$ and $\ensuremath{Y^\prime}\xspace(r)$ fare somewhat worse in reproducing the observed spacing, predicting an larger spacing 1.62 to 1.73 h.}
Nonetheless, Extended Data Fig.~\ref{fig.period_spacing} demonstrates that even independent of an absolute frequency scale, the period spacing of the $m=-2$ ring seismology spectrum strongly favors a modest stratification $N_{\rm max}/\omega_{\rm dyn}\sim2$ and hence an extended core-envelope interface.

\noindent{\bf Identifying g~mode order.}
With the spacing between W76.44 and W87.19/Maxwell accounted for as a single step in radial order $n$, there is a lingering possibility that the modes responsible for these waves are offset in radial order compared to our interpretation.
Indeed, an earlier seismic model\cite{2014Icar..242..283F} (F14) identified W87.19 with the mode \ensuremath{_{\,\,\,\,2}^{-2}g_3}\xspace, placing a strong resonance corresponding to \ensuremath{_{\,\,\,\,2}^{-2}g_2}\xspace at $\Omega_{\rm p}\approx2350\ {\rm deg\ d}^{-1}$, just inside the inner boundary of the C ring (see Fig. 5 of that work).
With W76.44 now detected in the inner C ring, it is reasonable to expect that with modest changes the same model could reproduce these two resonances with $n=2,\,3$ g~modes as in F14, in contrast with $n=1,\,2$ as in our favored model.
The $n=2,\,3$ interpretation would substantially change our conclusions regarding interior structure, generally requiring larger values of $N$ in the g~mode cavity and consequently a more abrupt composition gradient.

However, no such class of solutions emerged in the course of our democratic sampling algorithm, even in the most general cases.
{We furthermore find that samples mandating that W76.44 be produced by \ensuremath{_{\,\,\,\,2}^{-2}g_2}\xspace yield poor fits to both the seismology and gravity constraints.
The best models obtained in this way have a stable stratification from $\ensuremath{r_{\rm in}}\xspace/R_{\rm S}\approx0.1$ to $\ensuremath{r_{\rm out}}\xspace/R_{\rm S}\approx0.4$ but systematically underestimate the frequencies of W84.64 and W87.19/Maxwell by $\gtrsim100$ deg d$^{-1}$ and overestimate the magnitude of $J_4$ and $J_6$ by $\sim20$ and $\sim2.5$ ppm respectively.
}
We conclude that this scenario is not viable given the gravity and seismology constraints together, and hence that the $n=1,\,2$ g~modes are overwhelmingly the best interpretation for W76.44 and W87.19/Maxwell respectively.

\noindent{\bf Constraints at higher azimuthal order.}
Although our analysis focuses on $m=-2$ where W76.44 offers a stringent constraint on deep interior structure, ring seismology has probed prograde Saturn modes assuming all azimuthal orders from $m=-2$ through $-11$.
In addition to 4 bending waves~\cite{2019Icar..319..599F} at outer vertical resonances, the C ring contains 17 known density waves at outer Lindblad resonances\cite{2013AJ....146...12H,2014MNRAS.444.1369H,2016Icar..279...62F,2019Icar..319..599F,2019AJ....157...18H} which we compare to the modes of our preferred Saturn model in Extended Data Fig.~\ref{fig.m_rl}.

Beyond the lowest azimuthal orders $|m|\leq3$ already discussed, the observed waves are readily identifiable with Saturn modes that are essentially pure f~modes, consistent with expectations from simpler interior models\cite{1993Icar..106..508M,2019ApJ...871....1M}.
This reflects the fact that the f and g~modes naturally decouple as the angular degree $l$ is increased: f~modes are trapped increasingly close to the planetary surface while g~modes are trapped strongly in the deep region where $N>0$, and the coupling between the two is weakened by this diminishing overlap in their radial eigenfunctions.
Coupling is simultaneously weakened by their diverging frequencies: f~mode frequencies increase with $l$ as $\omega_f/\omega_{\rm dyn}\approx l^{1/2}$ whereas g~modes are confined to frequencies $\omega_g\lesssim N$. As a result their frequencies are dissimilar beyond $l\approx (N/\omega_{\rm dyn})^2$, which for the typical value $N_{\rm max}/\omega_{\rm dyn}=2$ obtained in our models implies that the f~modes cannot be strongly contaminated beyond $l\approx4$, consistent with the lack of observed g~modes starting at $m=-4$.

Indeed, near this decoupling threshold the modes $_{\,\,\,5}^{-5}f$ and $_{\,\,\,5}^{-5}g_1$ are near an avoided crossing and do exhibit some coupling, but the low level of overlap in their radial wavefunctions fails to enhance the surface amplitude of the g~mode to a level likely to be expressed in the rings.
Our preferred model does predict potentially observable $m=-3$ and $m=-4$ resonances in the inner C ring corresponding to the modes $_{\,\,\,3}^{-3}g_1$ and $_{\,\,\,4}^{-4}g_1$, a continuation of the same $n=1$ g~mode sequence of which we argue W76.44 is a part.
Another $m=-3$ resonance corresponding to $_{\,\,\,3}^{-3}g_2$ is predicted to fall near 86,000 km, but at an estimated wave amplitude $d\tau\approx3\%$ its detectability is more dubious.

Finally this model systematically overestimates the frequencies of the sectoral f~modes $_{\,\,\,l}^{-l}f$ with $l\geq4$, in contrast to the good agreement found in a previous study that applied a more simplistic three-layer structure model and neglected mode-mode coupling\cite{2019ApJ...871....1M}.
These modes are insensitive to deep structure, their frequencies instead dictated mostly by outer envelope structure and Saturn's rotation.
An adequate fit to the full set of ring seismology constraints therefore motivates more complex envelope structures than have been considered so far, as well as a treatment of realistic differential rotation profiles\cite{2019GeoRL..46..616G}.
Independent of these physical considerations, the results here also neglect third-order effects such as Coriolis-ellipticity coupling that may reduce mode frequencies by as much as a few percent, significant compared to the mismatch seen in Extended Data Fig.~\ref{fig.m_rl}.

\noindent{\bf Estimates of mode amplitudes.}
The mode excitation and dissipation processes operating in gas giants are unknown\cite{2018Icar..306..200M,2019ApJ...881..142W}.
The only amplitude constraints available for Saturn are the measured amplitudes of waves at Lindblad resonances with Saturn's sectoral $l=2-10$ f~modes\cite{2019AJ....157...18H}, and even this limited set evinces an unexpectedly complicated power spectrum.
To quantify the notion of detectability for the normal modes that we calculate, we simply assume that the relative mode amplitudes are given by energy equipartition and then anchor the overall spectrum such that a normal mode with frequency $\omega/\omega_{\rm dyn}=1$ receives a nondimensional mode amplitude $A_1=6\times10^{-10}$, with modes normalized by inertia \cite{2014Icar..242..283F}.
The ring response near a Lindblad resonance as a function of the effective forcing potential was computed by Goldreich \& Tremaine\cite{1979ApJ...233..857G} and described in detail for the special case of Saturn ring seismology by Fuller et al.\cite{2014Icar..231...34F} in their Section 5 and Appendix D.
This procedure yields optical depth semi-amplitudes of order 10\% near Lindblad resonances for the $l=2$, $m=-2$ modes in the C ring, roughly in line with the measured semi-amplitudes of the four $m=-2$ waves there\cite{2016Icar..279...62F,2014MNRAS.444.1369H,2019Icar..319..599F}.
These amplitudes are only an approximate guide due to Saturn's unknown mode excitation mechanism and our forgoing of any detailed local treatment for the mass density and opacity throughout the ring.

\noindent{\bf Influence of differential rotation on mode frequencies.}
{
We estimate the effect of deep zonal flows on mode frequencies by fitting the symmetric part of Saturn's observed cloud-level winds\cite{2011Icar..215...62G} with a polynomial and assuming that they decay with spherical radius following a logistic function:
}
\begin{align}
  \label{eq.rotation_law}
  \Omega(r,\mu) &= \Omega_{\rm S}\left[1+\eta(r)\sum_{s=0}^Sf_s\mu^{2s}\right]; \\
  \eta(r) &= \left(1+\exp\left[-\displaystyle\frac{2(r-r_0)}{\Delta\cdot R_{\rm S}}\right]\right)^{-1}.
\end{align}
{Here $\mu=\cos\theta$, $\Omega_{\rm S}=2\pi/(10.561\ {\rm h})$ is the rotation frequency assumed for the rigid deep interior, and the coefficients $f_s$ are obtained by projecting the observed winds\cite{2011Icar..215...62G} onto even Legendre polynomials and collecting terms of common order in $\mu$.
Taking $r_0=0.875\,R_{\rm S}$ and $\Delta=0.02$ yields a radial decay profile similar to that inferred from gravity and magnetic constraints by Galanti \& Kaspi\cite{2021MNRAS.501.2352G}; the resulting profile is shown in Extended Data Fig.~\ref{fig.firstord_dr}a-b for two choices of the latitudinal expansion order $S=0$ and $S=14$.}

{
For these rotation laws we calculate the first-order Coriolis perturbation to mode frequencies using inner products given by Dziembowski \& Goode\cite{1992ApJ...394..670D}.
This calculation does not account for avoided crossings induced by the differential rotation, a process that would affect the fine splitting observed between Maxwell/W87.19 and the three $m=-3$ patterns but not the three well-separated $m=-2$ frequencies that drive our main results.
}

{
Extended Data Fig.~\ref{fig.firstord_dr}c shows these corrections for sectoral ($l=-m$) modes in a typical interior model, in units of pattern speed.
Both the realistic latitudinal expansion $S=14$ and the trivial expansion $S=0$ produce similar frequency shifts: to first order, the modes experience the zonal flows as a super-rotating near-surface region.
For either rotation law, frequencies are relatively unchanged for modes of low degree such as $l=2$ because these mode eigenfunctions mostly occupy the inner regions of the planet (Figs.~\ref{fig.eigs_best}-\ref{fig.eigs_cavity_width}) where rigid rotation holds.
The wind contributions for all $\ell=2$ modes are less than 1 deg d$^{-1}$, an insignificant change compared to the conservative pattern speed uncertainty enforced in our likelihood function (of order $50\ {\rm deg\ d}^{-1}$), and compared to the residuals seen in Fig.~\ref{fig.ompat_j2n} for our best model. Differential rotation corrections are even smaller for g modes primarily trapped in the deep interior, but they are larger ($\sim \! 10\ {\rm deg\ d}^{-1}$) for high-$l$ f modes trapped close to the surface.
We conclude that Saturn's deep zonal flows do not affect our main results, but future work should address their contributions to the fine splitting at $m=-2,-3$, to the frequencies of high-degree f~modes, and to the differentially rotating background structure.
}



\end{document}